%% file: Main.tex
\documentclass[a4paper,12pt]{article}
\usepackage{jheppub}
\usepackage[usenames,dvipsnames,table]{xcolor}
\usepackage{graphicx,amsmath,amssymb,multirow,array,bm,mathrsfs,bbold,bbm}
\usepackage{appendix}
\usepackage{footmisc}
\usepackage{epsf,amsfonts,slashed,soul}
\usepackage{lipsum}
\usepackage{subfig}
\usepackage{pifont}
\usepackage{bbold}
\usepackage{dsfont}
\usepackage{tikz}
\usepackage{verbatim}
\usepackage{placeins}
\usepackage{physics}
\usepackage{graphicx}
\usepackage{comment}
\usepackage{xcolor}
\usepackage{mathtools}
\input{macros}
\usepackage{tikz} 
\usepackage[compat=1.1.0]{tikz-feynman}
\usepackage{feynmf}

\begin{document}

\title{Islands and traversable wormholes}

\author[a]{Ricardo Esp\'indola,} 
\author[b]{Viktor Jahnke}
\author[c,d]{and Keun-Young Kim} 
\affiliation[a]{Institute for Advanced Study, Tsinghua University, Beijing 100084, China.}
\affiliation[b]{Instituto de Física Teórica, UNESP-Universidade Estadual Paulista
R. Dr. Bento T. Ferraz 271, Bl. II, Sao Paulo 01140-070, SP, Brazil}

\affiliation[c]{Department of Physics and Photon Science, Gwangju Institute of Science and Technology,\\123 Cheomdan-gwagiro, Gwangju 61005, Korea}
\affiliation[d]{Research Center for Photon Science Technology, Gwangju Institute of Science and Technology, 123 Cheomdan-gwagiro, Gwangju 61005, Korea}

\emailAdd{ricardo.esro1@gmail.com}
\emailAdd{v.jahnke@unesp.br}
\emailAdd{fortoe@gist.ac.kr}

\abstract{We study information recovery in black hole evaporation using traversable wormhole protocols in AdS$_2$ Jackiw–Teitelboim gravity with matter fields coupled to an external bath. By introducing a simple non-local interaction between left and right radiation regions, we generate negative-energy shockwaves that render the wormhole traversable. We compute the resulting shifts in Kruskal coordinates, changes in the stress tensor, and backreaction effects on quantum extremal surfaces, finding a consistent decrease of the generalized entropy that aligns with the flat Page curve. Finally, we analyze the impact of negative-energy shocks on boundary two-point functions, providing a microscopic probe of island/radiation duality and discussing possible implications for experimental realizations in analog quantum-mechanical models.
}
\maketitle

\section{Introduction and Results}
Recent years have seen renewed interest in the black hole information paradox~\cite{InfoParadoxHawking}. In the 70s, Hawking demonstrated that black holes are not truly black when quantum effects are taken into account; instead, they emit radiation and gradually evaporate. 
However, at late times, the entropy of Hawking radiation exceeds the Bekenstein-Hawking entropy, suggesting a potential loss of information—a conclusion that clashes with unitarity in quantum mechanics. In the modern approach, this paradox is resolved by incorporating the entropy contribution of disconnected spacetime regions: so-called `islands.' These novel gravitating regions modify Hawking's original calculation by introducing new gravitational saddles; replica wormholes that dominate after the Page time. These solutions ultimately restore unitarity and recover the expected black hole Page curve \cite{Almheiri:2019psf, Penington:2019npb, Almheiri:2019hni, Almheiri:2019qdq, Penington:2019kki,Almheiri:2020cfm}.

Central to the modern information paradox resolution is the island formula, which allows to compute the fine-grained entropy of Hawking radiation semiclassically. Islands arise as result of extremizing the generalized entropy
\be\label{eq:island}
S(\mathbf{R}) ={\text{min}}~ \underset{I}{\text{ext}} \Big[ \frac{{A}(\partial I)}{4G_N} +  S_{\text{mat}}(R \cup I)\Big] \,,
\ee
where bold letters such as $\mathbf{R}$ represent quantities in the microscopic theory. In the semiclassical theory, $R$ represents the asymptotic region collecting Hawking radiation modes, $\partial I$ is the the boundary of the candidate island region $I$ with area $A(\partial I)$, and $S_{\text{mat}}$ is the von Neumann entropy of quantum fields. This formula has been extensively studied in various settings, including, for example \cite{Espindola:2022fqb, Bousso:2023kdj}. 

Importantly, this formula reveals a profound island/radiation duality through entanglement wedge reconstruction \cite{Almheiri:2019hni}: \emph{the quantum state in the gravitating island is encoded in the non-gravitating Hawking radiation}. In other words, the island acts as the \emph{hologram} of Hawking radiation. This novel realization suggests that, in principle, bulk/boundary reconstruction methods learned in AdS/CFT may also be used to reconstruct the black hole interior. 

A key question underlying the island/radiation duality follows: how can information in the island be operationally recovered from the radiation system? In this paper, we propose a possible answer to this question based on traversable wormholes inspired by Gao-Jafferis-Wall construction~\cite{Gao:2016bin} which
geometrically realize this retrieval and directly probe the island/radiation duality. The advantage of using geometric teleportation protocols, compared to other approaches such as the Petz map \cite{Penington:2019kki} or modular flow \cite{Chen:2019iro}, is that they employ simple operators and have been successfully applied in higher dimensional settings in AdS \cite{Freivogel:2019whb, Ahn:2020csv, Bintanja:2021xfs}, asymptotically flat spacetimes \cite{Maldacena:2018gjk,Fu:2019vco} and cosmological models \cite{Aalsma:2020aib, Aguilar-Gutierrez:2023ymx}.

In order to study black hole evaporation in a control setting, we revisit the model of AdS$_2$ Jackiw-Teitelboim (JT) gravity with conformal matter coupled to a rigid conformal field theory (CFT) \cite{Almheiri:2019qdq, Almheiri:2019yqk}. In Section \ref{sec:review}, we review important details of this model and set notation for the rest of the paper. In particular, we highlight the use of the global vacuum $\ket{\omega}$ where the stress tensor vanishes $T_{w^{\pm}w^{\pm}}=0$. Importantly, this theory contains a two-sided black hole that produces Hawking radiation. As a consequence of adding the entropy contribution of island regions, the entropy of Hawking radiation follows a `flat' Page curve at late times, where an emergent island covers a big part of the black hole interior. 

The two-sided black hole set up is `dual' to a couple of quantum mechanical dots connected with semi-infinite wires \cite{Almheiri:2019yqk} (see also \cite{Kruthoff:2021vgv}). In Section \ref{sec3}, we show a simple way of generating negative energy. Importantly, this mechanism do not need to assume holography. We thus proceed to turn-on a non-local coupling between fields in the microscopic theory by adding an interaction term to the action of the form
\be
S_{int} =  g \,\varphi_{\bf L} \varphi_{\bf R}~.
\ee
Here, $g$ is the coupling constant between $\varphi_{\bf L}$ and $\varphi_{\bf R}$, which represent matter fields in the entanglement wedge of the corresponding left and right radiation regions. These operator insertions excite the global state of the original joint system 
\be
\ket{\psi} = \exp({ig \varphi_{\bf L} \varphi_{\bf R}})\ket{\omega}~,
\ee 
which produce a change in the two-point functions between bulk fields $\phi$ in the semiclassical bulk theory. We compute the change in the stress tensor in two steps: first, we apply the two-point splitting method in (1+1)-flat spacetime; and second we map the result using a conformal transformation to the joint CFT plus bath system. It is worth noting that this procedure transcends holographic applications being a pure quantum field theory phenomenon. As consequence of the non-local interaction, a pair of shockwaves are produced featuring positive and negative energy for both the holomorphic $T_{w^+ w^+}$ and anti-holomorphic components $T_{w^- w^-}$.

In the semiclassical picture, negative energy shockwaves propagate into the gravitating region and positive energy shocks scape towards $\mathcal{I}^{\pm}$. In the gravity region negative shocks render the wormhole geometry traversable. By solving the equation of motion, we find the wormhole opening to have the form
\begin{equation}
\label{eq:shift}
    \Delta w^- = G_N \frac{\ell^2 \beta}{ \phi_r} \frac{g}{8\pi}\frac{1}{\omega^+_R} ~,
\end{equation}
with a similar expression for $\Delta w^+$. This wormhole shift allows for signals behind the horizon to scape towards $\mathcal{I}^{\pm}$. Crucially, this protocol represents a geometric mechanism for information information recovery from the island region by coupling the degrees of freedom in the radiation region $\mathbf{R}$. 

In Section \ref{sec4}, we provide evidence that this protocol is robust when considering backreaction effects on the island. Firstly, the negative energy shocks decrease the black hole mass according to the expression
\be
M(t) = M - E_s e^{-k(t-t_0)}~, \quad k=\frac{c}{24 \pi} \frac{8\pi G_N}{\phi_r}~,
\ee
where $E_s$ is the shockwave energy and $M$ is the original black hole mass. In JT gravity, the black hole mass is set by the Schwarzian action of the boundary particle $x(t)$. In the joint CFT plus bath system the flux across the gluing map determines the black hole mass change according to
\be
\frac{dM}{dt} = - \frac{d}{dt} \left( \frac{\phi_r}{8\pi G_N}  \{ x(t),t\} \right)  = T_{y^+ y^+} - T_{y^- y^-}~.
\ee
We solve the Shwarzian action in the presence of a negative energy shockwave (e.g. see equation (\ref{eq:gluingnegative})). It is important to stress the difference between this solution and the one with positive energy to model black hole evaporation \cite{Goto:2020wnk}. See Appendix \ref{app:A} for more details on the relationship between these solutions. 

One might be worried that backreaction will washed out the emergent island. Nevertheless, we show how quantum extremal surface changes in the presence of negative shocks. As a consequence, the generalized entropy decreases
\be
S = 2S_{BH} -  \frac{g (w_r^+ - w^+_R)w^+_R}{w^+_r(w^+_L-w^+_R)}~.
\ee
where $S_{BH}$ is the black hole entropy of the original black hole. At late times, the black hole thermalizes and the entropy follows a flat Page curve.

Finally, in Section \ref{sec5}, we study the effect of the negative energy shocks on the two-point correlation functions between operators in the joint system. From the microscopic point of view, these represent more realistic observables that interact with the environment. We finalize commenting on possible lines of research to explore in the future.

\paragraph{Organization} This paper is organized as follows. Section~\ref{sec:review} contains details of the  background material. In Section~\ref{sec3}, we show a simple way of generating negative energy with ordinary matter fields and without assuming holography. We couple two operators in flat space, derive the resulting stress-tensor from backreaction using the point splitting method, and map the result to the couple system of AdS plus bath by means of a conformal transformation. In Section~\ref{sec4}, we study the effect of negative energy backreaction. Specifically, we compute the change in the gluing map by solving the Schwarzian equation of motion. Then we show how the island changes in the presence of negative energy, evaluate the generalized entropy and show that at late times the black hole follows a flat Page curve. Finally, in Section~\ref{sec5} we study correlators in the dual quantum mechanical model. We conclude discussing future directions.

\section{Background material}
\label{sec:review}
In this section, we revisit a toy model for black hole evaporation involving JT gravity with conformal matter coupled to an external CFT \cite{Almheiri:2019yqk}. This framework provides a controlled setting to formulate a version of the black hole information paradox. We will be particularly interested in the setting involving a two-sided black hole in the global vacuum. 

\subsection{JT gravity coupled to a CFT}
Consider an AdS$_2$ black hole in JT gravity with conformal matter coupled to a rigid, non-dynamical bath. This system consists of JT gravity with matter given by a two-dimensional CFT with central charge $c$.  The action of the system has the form
\begin{align}
\label{eq:ActionJT}
I_{\rm JT + CFT} & = I_{\rm JT}[\phi,g,\varphi] + I_{\rm CFT} [g,\varphi] \\ \nonumber
  & = \frac{1}{16 \pi G_N}\int d^2 x \sqrt{-g} \left( \phi R  + 2 (\frac{\phi}{\ell} - \phi_0)\right) + I_{CFT}[g]~,
\end{align}
where $\phi_0$ is proportional to the extremal black hole entropy. At the AdS$_2$ boundary, we impose transparent boundary conditions, allowing quantum fluctuations to propagate into the non-gravitating bath. The matter CFT is taken to be identical across both the gravitating AdS$_2$ region and the rigid bath. 

At the boundary of the gravitating region ($\partial$AdS), we impose Dirichlet boundary conditions for the metric and dilaton
\be
\label{eq:JTbdycond.}
g_{uu}\Big\lvert_{\partial} = \frac{1}{\epsilon^2}~, ~ {\rm and} ~ \phi - \phi_0 \Big\lvert_{\partial} = \frac{\phi_r}{\epsilon}~,
\ee
where $u$ is the boundary proper time and $\epsilon$ the holographic radial cutoff in the direction orthogonal to $\partial$AdS. These conditions effectively freeze gravity at the boundary.

The equations of motion derived from the action (\ref{eq:ActionJT}) take the form
\be
\label{eq:eom}
R + 2 = 0~, \qquad \nabla_\mu \nabla_\nu \phi - g_{\mu \nu} \nabla^2 \phi + g_{\mu \nu} \phi = 8\pi G ~T_{\mu \nu}~.    
\ee
We are interested in a particular class of solutions of these equations, those that include two-sided black holes dual to a thermal CFT state. For convenience, we work in global coordinates $\omega^{\pm}$, which are related to Rindler coordinates by (see Appendix \ref{app:C} for more details on coordinate systems)
\be\label{eq:map}
w^{\pm} = \pm \ell e^{\pm 2\pi y_R^{\pm}/\beta}~, ~~ w^{\pm} = \mp \ell e^{\mp 2\pi y_L^{\pm}/\beta}~.
\ee
In these coordinates, the state of quantum fields is the vacuum CFT state $\ket{\omega}$, satisfying
\be
\langle T_{\pm \pm} \rangle= \frac{\pi c}{12 \beta^2}~ \xrightarrow[\text{}]{w^{\pm}(y^{\pm})} \langle T_{\pm \pm} \rangle_{\omega} =0~.
\ee
The equations of motion (\ref{eq:eom}) have the following solution for the metrics in the respective regions
\be\label{eq:metricw}
ds^2_{AdS} =- \frac{4\ell^4  dw^+ dw^-}{(\ell^2+w^+w^-)^2}~, ~~ ds^2_{CFT} = - \frac{\ell^2}{\epsilon^2} \left(\frac{\beta}{2\pi} \right)^2 \frac{dw^+ dw^-}{- w^+ w^-}~,
\ee
where $\beta$ is the black hole temperature, and $\epsilon$ is the UV cutoff ensuring agreement between the metrics at the timelike boundary $\partial$AdS.

The dilaton solution takes the simple form
\be
\phi(w^+,w^-) = \phi_0 + \frac{2 \pi \phi_r}{\beta} \frac{\ell^2- w^+ w^-}{\ell^2 + w^+ w^-}~.
\ee
While the gravitating region is locally AdS$_2$, the causal structure of spacetime is determined by the dilaton behavior. For instance, the spacetime singularity occurs where the dilaton vanishes
\be
    w^+ w^- = \ell^2 \frac{2 \pi \phi_r + \phi_0 \beta}{2 \pi \phi_r - \phi_0 \beta}~.
\ee

Although the system is in equilibrium, it exhibits a simplified version of the black hole information paradox. To analyze this, we compute the entanglement wedge of radiation applying the island formula, which reduces to extremizing the generalized entropy
\be
\label{eq:2Disland}
S(\mathbf{R}) =  \frac{{\phi}(\partial I)}{4G_N} +  S_{\text{mat}}(R \cup I)~,
\ee
where $\phi(\partial I)$ is the dilaton value (the area term) at the island boundary $\partial I$, and $S_{\rm mat}$ is the renormalized entropy of the quantum fields. The formula requires extremizing over all possible islands $I$ and minimize over all possible islands. 
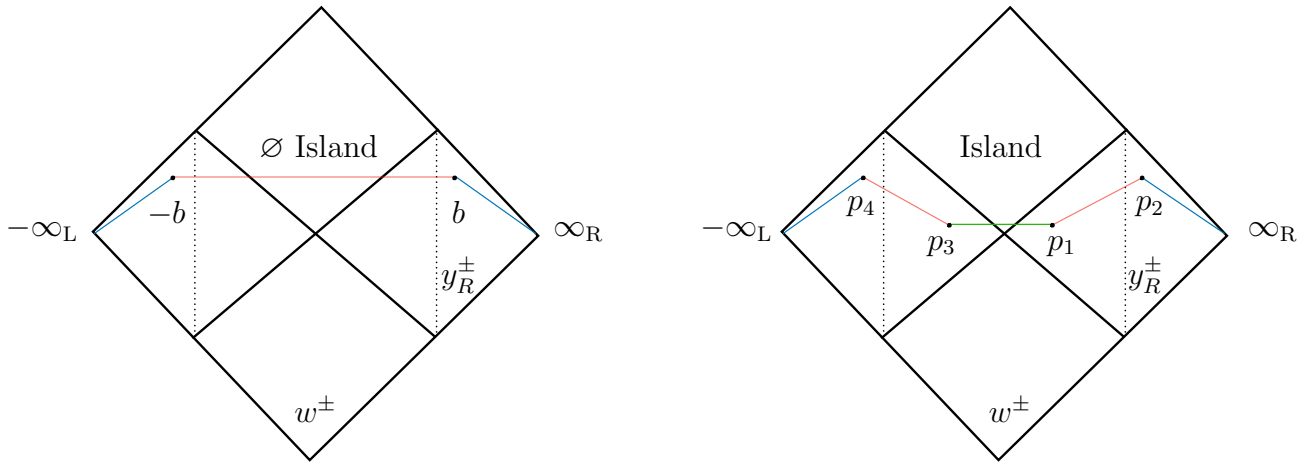
\begin{figure}[t]
 \centering
     \includegraphics[width=1.\linewidth]{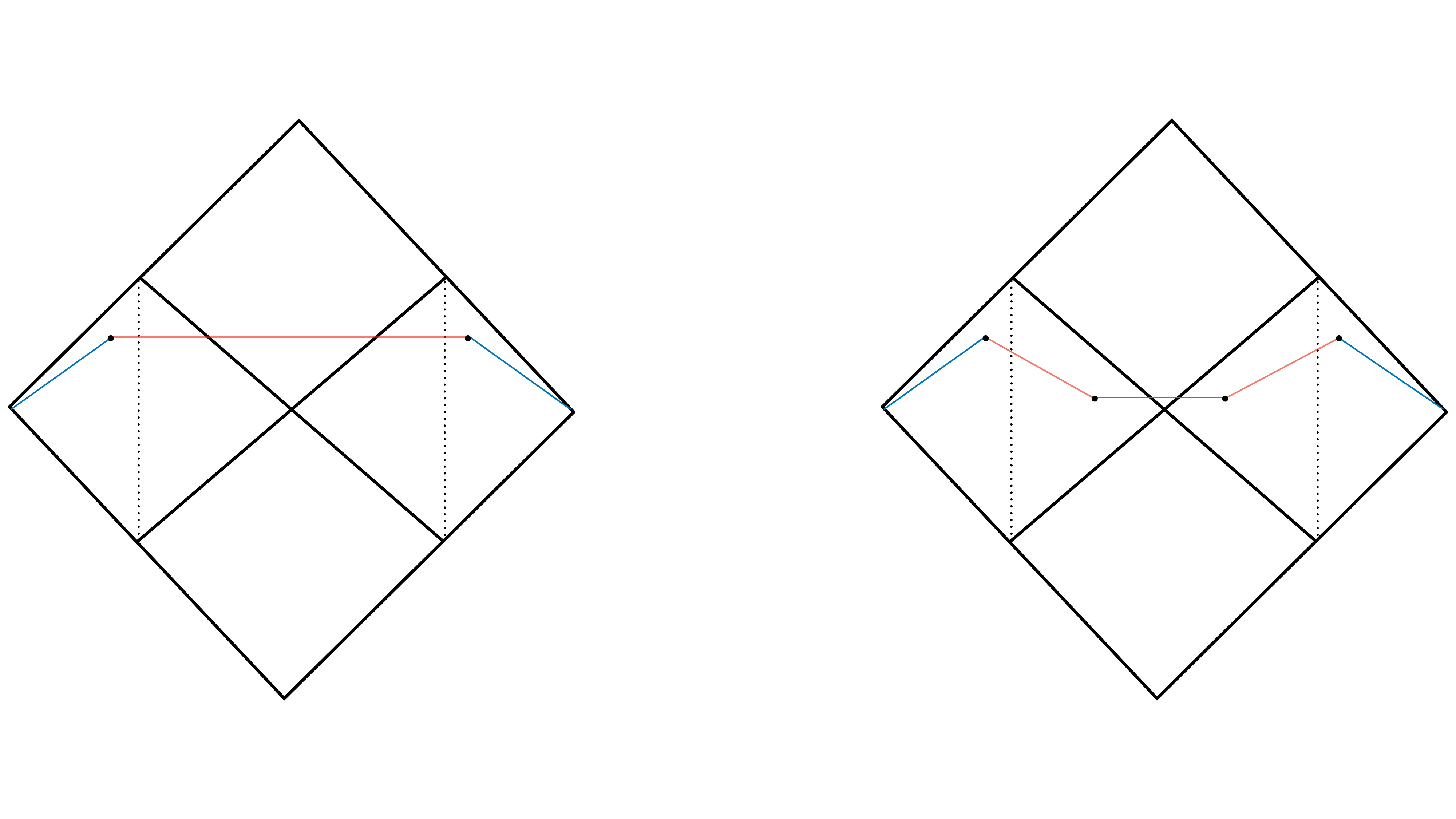}
    \put(-352,50){$w^\pm$}
    \put(-297,102){$y_R^\pm$}
    \put(-92,50){$w^\pm$}
    \put(-40,102){$y_R^\pm$}
    \put(-460,120){$-\infty_{\rm L}$}
    \put(-255,120){$\infty_{\rm R}$}
    \put(-200,120){$-\infty_{\rm L}$}
    \put(5,120){$\infty_{\rm R}$}
    \put(-145,130){$p_4$}
    \put(-115,115){$p_3$}
    \put(-70,115){$p_1$}
    \put(-37,130){$p_2$}
    \put(-103,150){{\rm Island}}    
     \put(-365,150){$\emptyset$ {\rm Island}}   
     \put(-407,125){$-b$}
     \put(-293,125){$b$}
     \vspace{-1cm}
     \caption{Eternal black hole coupled to a bath. Left: Early time configuration where the empty island dominates and gives rise to the linear increase of the entanglement entropy. Right: Configuration where an island shows up at late times. This saddle is responsible of the flatness of the entropy curve.}
\label{Fig:EternalBH}
\end{figure} 

We are interested in computing the fine grain entropy associated to the region ${\cal{R}}$, which consists of the union of intervals $[-b,0]$ and $[0,b]$. In semiclassical gravity, we evaluate the generalized entropy for the union of
intervals of radiation $R$ and candidate island $I$ depicted in Figure \ref{Fig:EternalBH}. At early times, the island is absent ($I=\emptyset$) so the entropy (\ref{eq:2Disland}) arises solely from the bulk matter contribution. Assuming the global state is pure on the non-compact Cauchy slice, we have $S_{\rm mat}(R \cup I) = S_{\rm mat}((R \cup I)^c)$%$S_{mat}\left[(-\infty, -b) \cup (b, \infty)\right] = S_{mat}\left[(-b,b)\right]$%

In the conformally flat metric
\be
ds^2 = -\frac{dw^+ dw^-}{\Omega^2(w^+,w^-)}~,
\ee
the matter entropy of an interval $(w_1,w_2)$ in the $w$-plane is given by the Cardy-Calabrese formula
\be
S =  \frac{c}{6} \ln \left( \frac{- \Delta w^+ \Delta w^-}{\epsilon_1 \epsilon_2 \Omega_1(w_1^+,w_1^-) \Omega_2(w_2^+, w_2^-)} \right)
\ee
where $\Delta w^\pm = w_2^\pm(y^\pm_2)-w_1^\pm(y^\pm_1)$ . The conformal factors at the points located in the bath region are
\be
\Omega_1(w_1^+,w_1^-) = \frac{2 \pi \ell}{\epsilon_{UV} \beta}\sqrt{-w_1^+ w_1^-}~ {\rm and} ~~ \Omega_2(w_2^+,w_2^-) = \frac{2 \pi \ell}{\epsilon_{UV} \beta}\sqrt{-w_2^+ w_2^-}~.
\ee
Substituting these in (\ref{eq:2Disland}), the entropy becomes
\be
S=\frac{c}{6} \ln \left( \left(\frac{\beta \epsilon_{UV}}{2 \pi \ell}\right)^2 \frac{-(w_2^+ -w_1^+)(w_2^- - w_1^-)}{\epsilon_1 \epsilon_2 \sqrt{w_2^+ w_2^- w_1^+ w_1^{-}}} \right) \,.
\ee
Applying the coordinate map (\ref{eq:map}), this simplifies to
\be
S = \frac{c}{3} \ln \frac{\beta}{\pi} \cosh \left(\frac{2\pi}{\beta}t \right) - {\rm UV~divergence}~.
\ee
The UV-divergent term, arising from flat-space short-distance effects, is regulator-dependent and can be ignored. At late times, we get the linear grow characteristic of the Hawking saddle. At late times ($t \gg \beta$),  the entropy grows linearly as
\be
S \sim \frac{c}{3} \frac{2\pi t}{\beta}~,
\ee
characteristic of the Hawking saddle. This result implies an information paradox: the entropy exceeds the Bekenstein-Hawking entropy of the eternal black hole,
\be
S_{BH} = \left(\phi_0 + \frac{2\pi \phi_r}{\beta}\right)~,
\ee
violating unitary evolution. 

We now consider an island $I = (p_3, p_1)$ in the gravitating region (see Figure \ref{Fig:EternalBH}). The entropy calculation reduces to evaluating the contribution from two intervals, $(p_4,p_3)$ and $(p_1,p_2)$. At late times, the points $p_2$ and $p_4$ approach infinity, the intervals become widely separated, and the entropy factorizes
\be 
S = 2\left(\frac{\phi(w_1^+,w_1^-)}{4 G_N} + \frac{c}{6} \ln\left(\frac{- \Delta w^+ \Delta w^-}{\epsilon_1 \epsilon_2 \Omega(w_1^+,w_1^-) \Omega(w_2^+, w_2^-)}\right) \right)~,
\ee
where UV-divergent terms are omitted. For simplicity, we fix the point $p_2$ at the boundary $b=0$, with coordinates $ (w_2^+, w_2^-)= (+ e^{2\pi t/\beta}, -e^{-2\pi t/\beta})$. The conformal factors at points $p_1$ (island) and $p_2$ are 
\be
\Omega(w_1^+,w_1^-) = \frac{1}{2 \ell^2}(\ell^2+ w_1^+ w_1^-)~, ~~ \Omega(w_2^+,w_2^-) = \frac{2 \pi \ell}{\beta \epsilon_{UV}} \sqrt{- w_2^+ w_2^-}~.
\ee
The generalized entropy (\ref{eq:Sgen1}) then becomes
\be\label{eq:Sgen1}
S_{\rm gen} = \frac{ \phi_0}{2 G_N} + \frac{4 \pi \phi_r}{ \beta G_N} \frac{1- w_1^+ w_1^-}{1+ w_1^+ w_1^-} + \frac{c}{3} \ln\left( \frac{\beta}{\pi} \right) + \frac{c}{3} \ln  \left( \frac{- \Delta w^+ \Delta w^-}{1+w_1^+ w_1^-}\right)~.
\ee
Extremizing $S_{\text{gen}}$ with respect to the point $(w_1^+,w_1^-)$ and taking the semiclassical limit $G_N \ll 1$, the location of the QES is
\be
\label{eq:QESsaddle}
\partial_{\pm} S_{\rm gen}=0 ~ \implies ~ \omega^{+}_I=-\frac{c\ell^2 \beta G_N}{24\pi \phi_r \omega_r^{-}},~~~\omega^{-}_I=-\frac{c\ell^2 \beta G_N}{24\pi \phi_r \omega_r^{+}}~.
\ee
Substituting these back into (\ref{eq:Sgen1}), the late-time entropy is
\begin{align}
S = 2 S_{BH}  - \frac{c^2\beta G_N}{72\pi \phi_r}+\frac{c}{3} \ln \left(\frac{2 \beta\epsilon_{UV}}{\pi\epsilon_1\epsilon_2}\right)~.
\end{align}
At late times, we then recover the Bekenstein-Hawking entropy, in units of $4\,G_N$. In the limit $\phi_r/c \gg \beta$, the entropy saturates at twice $S_{\text{BH}}$, reproducing the unitary Page curve via the island prescription.

\section{Non-local coupling}\label{sec3}
In this section, we propose a method for generating negative energy density in a $(1+1)$-dimensional CFT using simple operator insertions. In Section \ref{sec:3.1}, we show that introducing a non-local term to the action of the form 
$\delta S=g \varphi_L \varphi_R$
at a fixed time induces an expectation value of the stress tensor featuring both positive and negative energy shock waves. Section \ref{sec:3.2} maps this stress tensor result to the joint system of AdS with rigid bath, producing shockwaves that propagate into the gravitating regions. Finally, in Section \ref{sec:3.3}, we analyze the bulk response to the non-local coupling, showing that it renders the two-sided wormhole traversable and allows information in the island to escape. 
 
\subsection{First order calculation in flat spacetime}
\label{sec:3.1}
\begin{figure}[t]
\begin{small}
\begin{center}
\begin{tikzpicture}
\draw[thick, red] (0,5)--(2.5,2.5);
\draw[thick, cyan] (2.5,2.5)--(5,5);
\draw[thick, red] (5.5,2.5)--(8,5);
\draw[thick, cyan] (3,5)--(5.5,2.5);
\draw[<-,dashed] (0.5,6)--(4.5,2);
\node [above] at (0.5,6) {{$\boldsymbol{v}$}};
\draw[<-,dashed] (7.5,6)--(3.5,2);
\node [above] at (7.5,6) {{$\boldsymbol{u}$}}; 
\filldraw (2.5,2.5) circle (1.5pt) node[align=left, left,left] {\scriptsize $(u_{L},v_L)$};
\filldraw (5.5,2.5) circle (1.5pt) node[align=right, right] {\scriptsize $(u_{R},v_R)$};
\filldraw (4,2.5) circle (1.5pt) node[align=right, right] {\scriptsize $(0,0)$};
\end{tikzpicture}
\end{center}
\end{small}
\caption{The blue/red lines represent the regions of space where we have negative/positive energy density.}
\label{Fig:Coupling}
\end{figure}
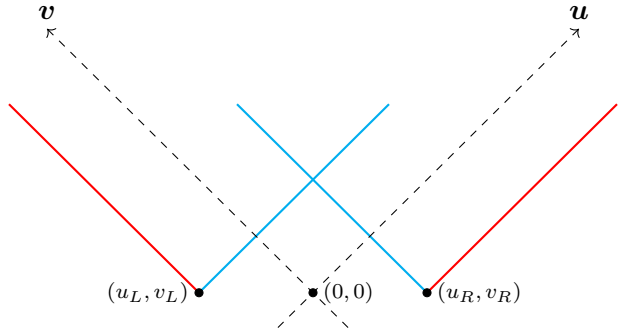

We consider a free massless scalar field propagating in flat space. For convenience, we adopt light-cone coordinates
\be
\label{eq:uv}
ds^2 = - du dv \quad , \quad u = t+x~, ~~ v = t-x ~,
\ee
where the action of the scalar field is given by
\begin{equation}
S=-\int \frac{1}{2}\partial_{\mu}\varphi\partial^{\mu}\varphi ~.
\end{equation}
We deform the system by introducing an interaction term at time $t=0$ of the form
\begin{equation}
\label{eq:coupling}
\delta S_{\rm int}=g \varphi_L \varphi_R~,
\end{equation}
where $\varphi_{L,R}$ denotes the field operator $\varphi$ evaluated at two spacelike-separated points $x_{L}$ and $x_{R}$ (see Fig. \ref{Fig:Coupling}).
This interaction excites the system, generating a perturbed state
\be
\ket{\Psi}=e^{ig\varphi_{L}\varphi_{R}}|\omega \rangle~,
\ee 
where $\ket{\omega}$ is the initial global vacuum state. We now compute the expectation value of the normal ordered stress-energy tensor on the perturbed state. The ingoing component evaluates to
\begin{align}
\label{Tuu}
\lel \Psi|:T_{uu}(u):|\Psi\rir &=
\lel e^{-ig\varphi_{L}\varphi_{R}}:\partial_{u}\varphi\partial_{u}\varphi:e^{ig\varphi_{L}\varphi_{R}}\rir \\ \nonumber
&=\lel \left(1-ig\varphi_{L}\varphi_{R}\right):\partial_{u}\varphi\partial_{u}\varphi:\left(1+ig\varphi_{L}\varphi_{R}\right)\rir \\ \nonumber
&= ig\lel :\partial_{u}\varphi\partial_{u}\varphi:\varphi_{L}\varphi_{R}\rir-ig\lel \varphi_{L}\varphi_{R}:\partial_{u}\varphi\partial_{u}\varphi:\rir \\ \nonumber 
&=ig\lel\left[:\partial_{u}\varphi\partial_{u}\varphi:,\varphi_{L}\varphi_{R}\right]\rir,
\end{align}
where we expanded to linear order in $g$ and used the vanishing of $\langle:\partial_{u}\varphi\partial_{u}\varphi: \rangle$ in the vacuum. Defining the correlation function
\be
C\equiv \lel :\partial_{u}\varphi\partial_{u}\varphi:\varphi_{L}\varphi_{R}\rir~,
\ee
and assuming $\varphi_{R},\varphi_{L}$ and $\partial_{u}\varphi$ are Hermitian, the commutator reduces to
\be
\lel\left[:\partial_{u}\varphi\partial_{u}\varphi:,\varphi_{L}\varphi_{R}\right]\rir = - 2 \Im~C~.
\ee
This correlator quantifies wormhole traversability in the context of gravity (see Appendix \ref{app:B} for more details). Thus, we obtain
\begin{equation}
\lel \Psi|:T_{uu}(u):|\Psi\rir=-2g\text{Im}\left(\lel :\partial_{u}\varphi\partial_{u}\varphi:\varphi_{L}\varphi_{R}\rir\right)=-4g\text{Im}\left(\lel \partial_{u}\varphi\varphi_{L}\rir\lel \partial_{u}\varphi\varphi_{R}\rir\right),
\end{equation}
where the second equality follows from Wick contractions. Since the correlators of interest are not time-ordered, we evaluate them using the Wightman function. The Wightman function for the free massless scalar in two dimensions is \cite{Afshordi:2012ez}
\begin{equation}
\begin{split} 
W(t,x;t',x')=&\lel \varphi(t,x)\varphi(t',x')\rir=-\frac{1}{4\pi}\left[\log{\left[i\mu\left(\Delta t+\Delta x -i\epsilon\right)\right]}+\log{\left[i\mu\left(\Delta t-\Delta x -i\epsilon\right)\right]}\right],
\end{split}
\end{equation}
where $\mu$ is an infrared cutoff. In light-cone coordinates (\ref{eq:uv}), this becomes 
\be
\label{wightmanlight}
W(u,v;u',v')=\lel\varphi(u,v)\varphi(u',v')\rir = -\frac{1}{4\pi} \left( \log{\left[i\mu\left(\Delta u -i\epsilon\right)\right]}+\log{\left[i\mu\left(\Delta v -i\epsilon\right)\right]} \right) .
\ee
Using \eqref{wightmanlight} and taking $\epsilon \rightarrow 0$, we compute \eqref{Tuu}
\begin{equation}
\label{Tuuf}
\lel \Psi|:T_{uu}(u):|\Psi\rir=-\frac{g}{4\pi}\left(\frac{\delta(u-u_{R})}{u-u_L}+\frac{\delta(u-u_{L})}{u-u_R}\right)\theta \left(\frac{u+v}{2}\right)
\end{equation}
where we employed the identity
\begin{equation}
\label{deltarep}
\delta(x)=\frac{1}{\pi}\lim_{\epsilon\rightarrow 0}\frac{\epsilon}{x^2+\epsilon^2}.
\end{equation}
The Heaviside step function $\theta \left(\frac{u+v}{2}\right)=\theta(t)$ ensures causality (non-zero perturbation only present for $t>0$).

An similar calculation yields the component $\lel :T_{vv}(v):\rir$
\begin{equation}
\label{Tvv}
\lel \Psi|:T_{vv}(v):|\Psi\rir=-\frac{g}{4\pi}\left(\frac{\delta(v-v_{R})}{v-v_L}+\frac{\delta(v-v_{L})}{v-v_R}\right) \theta\left(\frac{u+v}{2}\right).
\end{equation}

A light ray traveling along $v=0$ will only ``pass through" negative energy density
\begin{equation}
\begin{split}
    \int_{-\infty}^{\infty} du \, \langle T_{uu} \rangle &=-\frac{g}{4\pi}\int_{-\infty}^{\infty} du\, \left(\frac{\delta(u-u_{R})}{u-u_L}+\frac{\delta(u-u_{L})}{u-u_R}\right)\theta \left(\frac{u+v}{2}\right)\Big|_{v=0}\\
    &=-\frac{g}{4\pi}\int_{0}^{\infty} du\, \left( \frac{\delta(u-u_{R})}{u-u_L} +\frac{\delta(u-u_L)}{u-u_R}\right)=-\frac{g}{4\pi}\frac{1}{u_R-u_L} <0\,,
\end{split}
\end{equation}
where we used that $u_L<0$ and $u_R>0$. Conversely, a light ray traveling along $u=0$ intersects positive energy resulting in $\int_{-\infty}^{\infty} \langle T_{vv}\rangle dv<0$.

\subsection{Conformal transformation}
\label{sec:3.2}
The combined system of two-sided black hole coupled to the bath is described in terms of the conformally flat metric
\be
ds^2 = -\frac{dw^+ dw^-}{\Omega^2(w^+,w^-)}~. 
\ee
So far we have discussed how to generate negative energy in in Minkowski spacetime. In order to incorporate gravity to the picture, we need to take our stress tensor components (\ref{Tuuf}) and (\ref{Tvv}), and map them to the conformally flat metric. The stress tensor transformation is given by the conformal anomaly 
\be
Z[g = e^{2\omega} \hat{g}]= \exp{i\frac{c}{24\pi}\int d^2x \sqrt{\hat{g}} \Big( \hat{R}\omega + (\hat{\nabla} \omega)^2\Big)} Z[\hat{g}]~,
\ee
where
\be
T_{\mu \nu}^g = T_{\mu \nu}^\eta - \frac{c}{12 \pi} \left[ \frac{1}{2} \eta_{\mu \nu} \frac{(\partial \Omega)^2}{\Omega^2} + \frac{\partial_{\mu} \partial_\nu \Omega}{\Omega} - \eta_{\mu \nu} \frac{\partial^2 \Omega}{\Omega} \right]~, ~~~ \Omega = e^{- \omega}~.
\ee
The stress tensor components in the conformally flat metric have the form
\begin{align}
\label{eq:TinGR}
T_{w^+ w^+} &= T_{++}^\eta - \frac{c}{12 \pi} \frac{\partial_+^2 \Omega}{\Omega^2} \\
T_{w^-w^-} &= T_{--}^\eta - \frac{c}{12 \pi} \frac{\partial_-^2 \Omega}{\Omega^2}~.
\end{align}
We are interested on the stress tensor along the AdS boundary $\omega^+ \omega^- = -1$. Hence, the perturbed stress tensor components are
\begin{align}\label{eq:Tnew}
T_{w^+ w^+}& = -\frac{g}{4\pi} \left( \frac{\delta(w^+ - w^+_R)}{w^+-w^+_L} + \frac{\delta(w^+ - w^+_L)}{w^+ - w^+_R} \right) \theta \left(\frac{w^++w^-}{2}\right)~,\\ \nonumber
T_{w^- w^-}&= -\frac{g}{4\pi}\left(\frac{\delta(w^- -w^-_{R})}{w^- - w^-_L}+\frac{\delta( w^- - w^-_{L})}{w^- -w^-_R}\right)\theta \left(\frac{w^++w^-}{2}\right)~.
\end{align}

\subsection{Wormhole opening}\label{sec:3.3}
\begin{figure}[t]
\centering
\includegraphics[width=.8\linewidth]{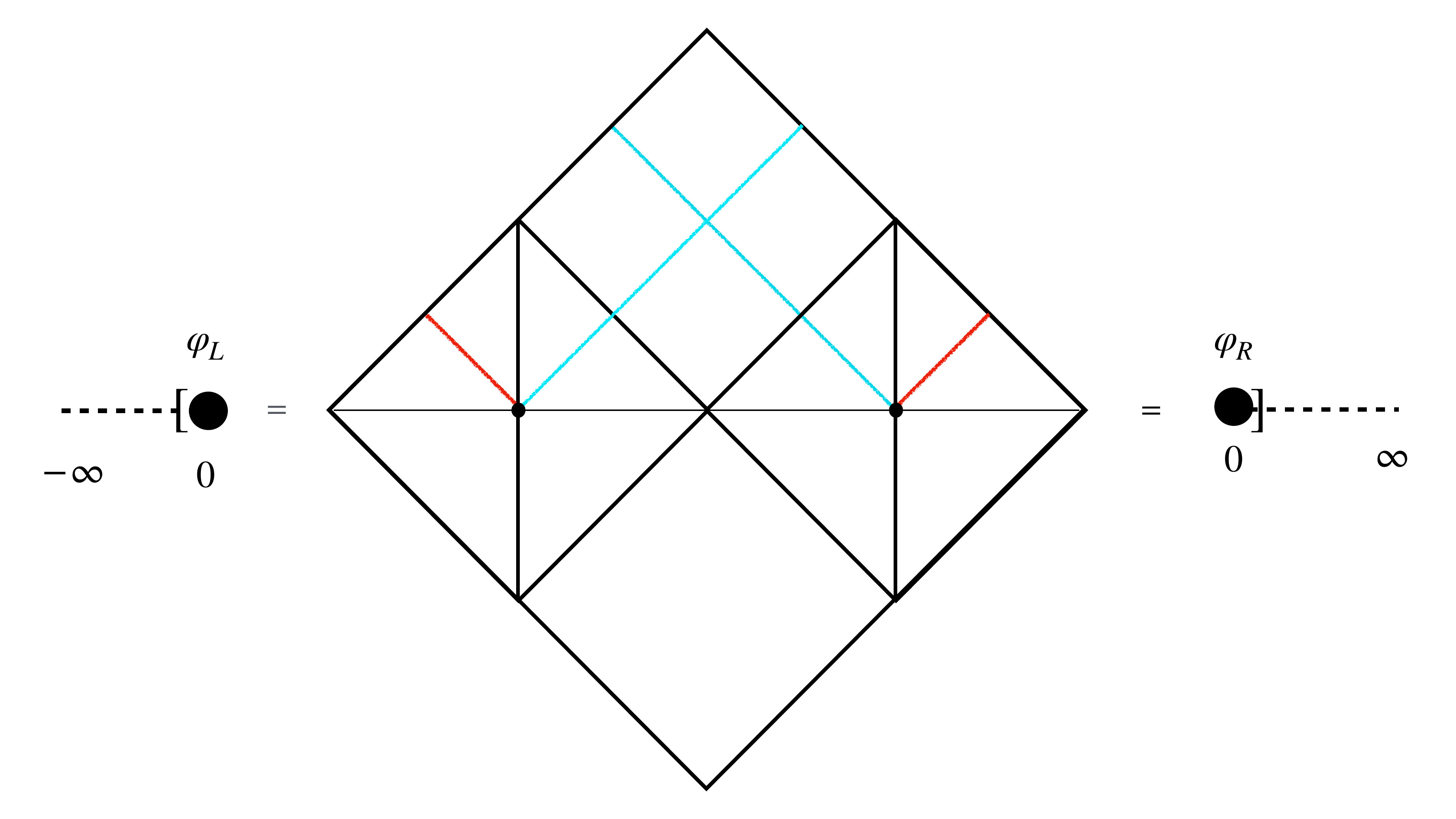}
     \put(-150,20){$w^\pm$}
    %\put(-140,102){$y_R^\pm$}
    %\put(-240,90){$\phi_L$}
     %\put(-130,90){$\phi_R$}
     \caption{Energy shock waves produced by the non-local deformation in the conformaly flat geometry.}
\label{Fig:EternalBH2}
\end{figure}

We now study the bulk geometry response to the non-local coupling, which generates two shock wave pairs: one propagating into the gravitating region and another escaping to infinity. Our focus will be on the pair of negative energy shocks entering the gravitating region, as these are responsible for rendering the wormhole traversable (see Fig. \ref{Fig:EternalBH2}). 

In JT gravity, the equations of motion (\ref{eq:eom}) impose the metric to be locally AdS$_2$, implying the existence of coordinates where backreaction effects become trivial. Specifically, there are two natural gauge choices available. The first corresponds to keeping the metric fixed while allowing the dilaton to be modified. The second choice amounts to modifying the metric while keeping the dilaton fixed \cite{Maldacena:2016upp}. In order to compute the shift produced by the non-local deformation, the latter gauge proves particularly convenient. We focus on the right exterior black hole due to symmetry. Thus the shock wave produced by $T_{++}$ will affect the bulk. In this gauge, the metric takes the form
\be
ds^2 = - \frac{4 \ell^4 dw^+ dw^-}{(\ell^2 + w^+ w^-)} + h_{++}(w^+) (dw^+)^2~,
\ee
while the dilaton maintains its non-perturbed form 
\be\label{eq:dilaton1}
\phi = \phi_0 + \frac{2 \pi \phi_r}{\beta} \frac{\ell^2- w^+ w^-}{\ell^2 + w^+ w^-}~.
\ee
Linearized Einstein's equations acquire the form 
\be
(++):  - \frac{2\pi \phi_r}{\beta \ell^2} \left(  h_{++} - \frac{1}{2} w^+ h_{++}' \right)  = 8 \pi G_N \lel T_{++}\rir~.
\ee
Integrating this equation with respect to $w^+$ and using the fact that the metric perturbation vanishes at infinity, we get
\be \label{eq-ANE-Tmn}
- \frac{2 \pi \phi_r}{\beta \ell^2} \int dw^+~ h_{++} = 8 \pi G_N \int dw^+ \lel T_{++}\rir~.
\ee
In a general perturbed background, very close to the horizon a null ray will have a shift of the form
\be\label{Eq:Shift}
\Delta V =- \frac{1}{2 g_{UV}(0)} \int dU h_{UU}~.
\ee
The shift is related to the stress tensor as follows
\begin{equation}
    \Delta w^- = -\frac{\beta \ell^2 G_N}{ \phi_r} \int dw^+ \lel T_{w^+ w^+}\rir~.
\end{equation}
We can now use the stress tensor profile produced by the non-local coupling (\ref{eq:TinGR}), and compute the integral
\begin{equation}
\label{Tuuf}
\int d\omega_{+} \lel T_{++}(\omega_{+})\rir=-\frac{g}{4\pi}\left(\frac{1}{\omega^+_R-\omega^+_L}\right)=-\frac{g}{4\pi}\frac{1}{2\omega^+_R},~\omega^+_R=-\omega^+_L~.
\end{equation}
Finally, the shift (\ref{Eq:Shift}) produced by the coupling (\ref{eq:coupling}) becomes
\begin{equation}
\label{eq:shift}
    \Delta w^- = G_N \frac{\ell^2 \beta}{ \phi_r} \frac{g}{8\pi}\frac{1}{\omega^+_R} ~.
\end{equation}
Importantly, this expression coincides perfectly with the estimation in \cite{Almheiri:2019yqk}.

It is instructive to compare the result in Eq.~\eqref{Eq:Shift} with the Gao–Jafferis–Wall expression~\cite{Gao:2016bin} (see Appendix~\ref{app:B} for a review of the derivation):  
\begin{equation} \label{eq-GJW} 
   \delta X^-=-g \, G_N \, \Delta \left( \frac{U_0}{1+U_0^2} \right)^{2\Delta+1}\,, 
   \qquad U_0=e^{\frac{2\pi}{\beta}t_0}\,,
\end{equation}  
where the Kruskal coordinate $U = e^{\tfrac{2\pi}{\beta} t}$
plays a role analogous to the coordinate 
\(\omega_R^+\) introduced above.\footnote{The expression in \eqref{eq-GJW} was derived with \(\ell=1\).}  
In both setups, the wormhole opening is proportional to \(g\,G_N\), reflecting the fact that we are working to linear order in the coupling \(g\) and employing point-splitting, with Einstein’s equations then relating the averaged null energy (ANE) to the expectation value of the stress tensor, cf.~\eqref{eq-ANE-Tmn}.  The dependence on the Kruskal coordinate differs in the two cases. This difference arises because the GJW result was derived for a massive bulk scalar field propagating in AdS$_2$, dual to a boundary operator of scaling dimension \(\Delta\). By contrast, our result~\eqref{eq:shift} follows from a massless scalar field , whose two-point function has a logarithmic form (cf.~Eq.~\eqref{wightmanlight}), whereas in the GJW analysis the relevant two-point function is conformal and fixed by the scaling dimension \(\Delta\). These results highlight the fundamental
distinction between GJW result using holography and our approach.

\section{Effects of the negative energy shockwave} \label{sec4} 
In this section, we investigate the effect of two negative energy shockwaves originating from the bath regions that cross the boundary at time $t_0$ with energy $E_\text{shock}<0$. We begin in Section \ref{sec:negative} by deriving the change in the dilaton solution. Specifically, we work in a gauge where the metric is fixed and we solve the Schwarzian equation of motion for the gluing map. In Section \ref{sec:islands}, we compute the entropy of Hawking radiation by extremizing the island formula, which yields distinct saddles corresponding to two scenarios.  one where the candidate island lies within the future lightcone of the shockwave (Section \ref{sec:inside}) and another where it lies outside (Section \ref{sec:outside}). The latter saddle is responsible for recovering the Page curve of the perturbed black hole. 

\subsection{Negative-energy solution}
\label{sec:negative}
The general solution to the dilaton equation of motion (\ref{eq:eom}) has the form \cite{Almheiri:2014cka}
\be
\label{eq:phigeneral}
\phi(w^+,w^-) = \phi_0 + \frac{2 \pi \phi_r}{\beta} \frac{\ell^2- w^+ w^-}{\ell^2 + w^+ w^-} + \delta \phi^+(w^+,w^-) + \delta \phi^-(w^+,w^-)~, 
\ee
where the perturbations are described by 
\begin{align}
\label{eq:deltaphi}
\delta \phi^+(w^+,w^-) & = \frac{8 \pi G  \ell^2 w^-}{\ell^2 + w^+ w^-} \int_0^{w^+}~dx~ \left(\frac{w^+}{l} - \frac{x}{\ell} \right) \left(\frac{\ell}{w^-} + \frac{x}{\ell} \right) T_{++}(x)~, \\ \nonumber
\delta \phi^-(w^+,w^-)& = \frac{8 \pi G  \ell^2 w^+}{\ell^2 + w^+ w^-} \int_0^{w^-}~dx~ \left(\frac{w^-}{l} - \frac{x}{\ell} \right) \left(\frac{\ell}{w^+} + \frac{x}{\ell} \right) T_{--}(x) ~.
\end{align}
It is particularly useful to work in a gauge where the metric remains fixed while only the dilaton is perturbed. This choice will be particularly convenient for identifying island regions, as expressing the dilaton in terms of the gluing map between the rigid and gravitating systems greatly simplifies the extremization of generalized entropy. This naturally leads us to consider the Schwarzian action in JT gravity.

In this framework, the gravitational dynamics is encoded by the boundary particle trajectory $x(t)$, which equivalently serves as the gluing map when coupled to the bath. We focus on the right exterior of the black hole, where only one shock enters the gravitating region (see Fig. \ref{Fig:EternalBH2}); a similar analysis applies to the left exterior.

The ingoing shockwave produces two important effects. It increases the temperature of the black hole, and it modifies the dilaton from its original form. These modifications are captured by the dynamical gluing map, which can be determined from the Schwarzian action
\be
I_{JT} = - \frac{\phi_r}{8 \pi G_N} \int dt \{ x(t),t\} + {\rm topological}~.
\ee 
The resulting equation of motion encodes energy conservation
\be
\frac{dM}{dt} = - \frac{d}{dt} \left( \frac{\phi_r}{8\pi G_N}  \{ x(t),t\} \right)  = T_{y^+ y^+} - T_{y^- y^-}~,
\ee
where $M$ is the ADM mass. The non-local deformation begins at $t=t_0$, with the eternal black hole mass providing the initial condition for $t<t_0$
\be
M = \frac{\phi_r}{4 G_N} \frac{\pi}{\beta^2}~.
\ee 
In the right black hole exterior, only the ingoing shock  (set by $x(y^+)$) excites the state,  leaving the outgoing stress tensor component unaffected, and we have for both components
\be
\langle T_{++} \rangle= \frac{\pi c}{12 \beta^2}~ , ~~~ \langle T_{++} \rangle =- \frac{c}{24 \pi} \{ x(y^+), y^+\}~.
\ee
The Schwarzian equation of motion then acquires the form\footnote{We will use notation where $T_{y^+ y^+} = - E_s$ so that $E_s>0$.}
\be\label{eq:eomsch}
- \partial_t \left( \frac{\phi_r}{8\pi G_N}   \{ x(t),t\}  \right) = E_\psi \delta(t-t_0) + \frac{c}{24 \pi}  \{ x(t),t\}  + \frac{c \pi}{12 \beta^2}~.
\ee
Introducing the parameter  $k = \frac{c}{24 \pi} \frac{8 \pi G_N}{\phi_r}$, we rewrite this as
\be\label{eq:x(t)eq}
- \partial_t   \{ x(t),t\} =  -\frac{24 \pi k E_s}{c} \delta(t-t_0) + k   \{ x(t),t\} + k \frac{2\pi^2}{\beta^2} ~.
\ee
We propose an ansatz for the Schwarzian derivative
\be\label{ansatz}
\{ x, t\} =  A + B e^{- C t}~,
\ee
with integration constants fixed by substitution into (\ref{eq:x(t)eq})
\be
A = -  \frac{2 \pi^2}{\beta^2}~,~ C=k~.
\ee
The last integration constant, $B$, is determined by matching the eternal black hole solution at early times ($E(t_0) = M - E_s$), yielding
\be
B = \frac{24 \pi k }{c} E_{s} e^{k t_0}~. 
\ee
The mass at times $t>t_0$ can then be expressed as
\be
M(t) =-\frac{\phi_r}{8\pi G_N} \{x(t),t \}=-\frac{\phi_r}{8\pi G_N}\left( A + B e^{- C t}\right)= \frac{\phi_r}{4 G_N} \frac{\pi}{\beta^2}-E_s e^{-k(t-t_0)}\,,
\ee
which shows that the negative energy shock wave initially decreases the black hole mass by $E_s$. However, the contact with the bath makes the black hole's mass to restore its original value as $t \rightarrow \infty$.

Rewriting in terms of $\phi := \log x'(t)$, the ansatz becomes
\be
-\frac{1}{2}(\phi')^2 + \phi'' = A + B e^{-Ct}~.
\ee
We now introduce the variables
\be
F := e^{- \frac{1}{2}\phi}~~{\rm and}~~  y := \nu \sqrt{\frac{E_s}{M}} e^{-\frac{k}{2}(t-t_0)}~,
\ee
and get the Bessel equation
\be
\left(y^2 \frac{d^2}{dy^2} + y \frac{d}{dy}  + (y^2 + \nu^2)\right) F =0~, ~~ \nu = \frac{2\pi}{k \beta}~~~.
\ee
The solution is given in terms of Bessel functions of first and second kind respectively
\be
F(y) = c_1 J_\nu(y) + c_2 Y_\nu (y)~.
\ee
Finally, inverting this expression we get the gluing map, 
\be\label{eq:gluingnegative}
x(t) =  \int\limits_{t_0}^t dt' \frac{1}{\left[c_1 J_\nu(y(t')) + c_2 Y_\nu (y(t'))\right]^2} +c_3 ~.
\ee
This solution notably differs from the positive-energy shockwave result in \cite{Goto:2020wnk}. We refer to the interested reader to Appendix \ref{app:A}, where we analyze the positive energy solution in more detail.

We fix the constants in (\ref{eq:gluingnegative}) by imposing the matching conditions with the original (``vacuum") gluing map at time $t=t_0$
\be
x(t_0) = e^{\frac{2\pi}{\beta}t_0}~,~~ x'(t_0) =\frac{2\pi}{\beta} e^{\frac{2\pi}{\beta}t_0} , ~~ {\rm and}~~ x''(t_0)  = \left(\frac{2\pi}{\beta}\right)^2 e^{\frac{2\pi}{\beta}t_0} ~,
\ee
The gluing map can be expressed in terms of Bessel functions as 
\be
x(t) = e^{\frac{2\pi}{\beta}L}  \left(\frac{Y_{\nu}(-y(t)) J_{\nu-1}(-y_0 ) - J_\nu(-y(t)) Y_{\nu-1}(-y_0 )}{Y_{\nu+1}(-y_0 ) J_\nu(-y(t)) - J_{\nu +1}(-y_0) Y_{\nu}(-y(t)} \right)~,
\ee
where we defined $y_0 := y(t_0) $.

At late times $t \gg t_0$, the solution asymptotes to
\be\label{eq:xlate}
x(t) \approx x_{\infty} - (x_\infty - x(t_0)) e^{- 2 \nu (\eta(y_0) - \eta(y))}~
\ee
with
\be
\eta(y) = \sqrt{1+y^2} + \log\left( \frac{y}{1+\sqrt{1+y^2}} \right)~, ~ {\rm and} ~ x_{\infty}:= x(t \rightarrow \infty) \approx -  e^{\frac{2\pi}{\beta}L} \frac{J_{\nu-1}(-y_0)}{J_{\nu +1}(-y_0)}~.
\ee
To analyze the horizon shift, we transform to the $w-$plane by means of the following change of coordinates
\be\label{eq:xwmap}
x^+ = w^+ ~ ,~ x^- = -1/w^-~.
\ee
In these coordinates, the horizon is located at the new location 
\be
w^-_{\rm hzn }=-\frac{1}{x_{\infty}} \approx \frac{g}{4 e^{\frac{2\pi t_0}{\beta}}} \frac{E_s}{M} - \frac{\beta}{8\pi} \frac{g k}{ e^{\frac{2\pi t_0}{\beta}}} \frac{E_s}{M} + \cO(g^2)~.
\ee
Here we redefined the negative energy as follows $E_s \rightarrow g E_s$. This corresponds to a horizon displacement
\be\label{eq:shift2}
\Delta w^- \approx \frac{g G_N}{ \pi \phi_r}  \frac{\beta^2 E_s}{e^{\frac{2\pi t_0}{\beta}}}~.
\ee
The order $\mathcal{O}(g)$ term agrees precisely with \cite{Almheiri:2019yqk}.
We see that the original black hole horizon recedes as expected after the backreaction provoked by the negative energy. A little diamond of flat space opens up in the middle of spacetime.

\subsection{Islands}
\label{sec:islands}
Having analyzed the backreaction effects on the dilaton solution and the resulting horizon shift (\ref{eq:shift2}), we now turn to the computation of the generalized entropy. As time progresses, the black hole emits Hawking radiation, which escapes toward $\mathcal{I^+}$ or falls back into the black hole. Beyond the Page time, a non-trivial island emerges. In what follows, we focus on studying the post-Page time regime and examine the black hole more closely. As we will demonstrate, distinct island configurations arise depending on whether the QES, $\partial I$, lies inside or outside the future lightcone of the shockwave. 

\subsubsection{$\partial I$ outside the shock wave future lightcone}
\label{sec:outside}
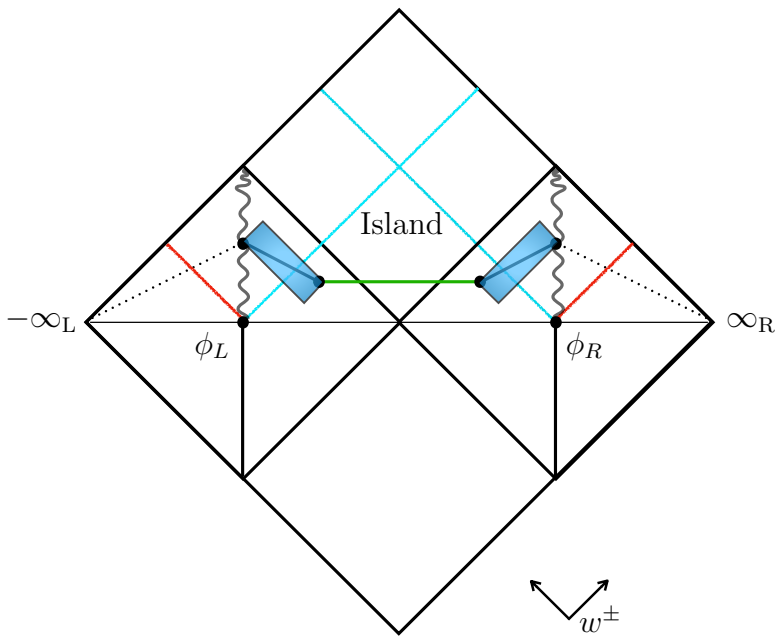
\begin{figure}[t]
 \centering
     \includegraphics[width=1.05\linewidth]{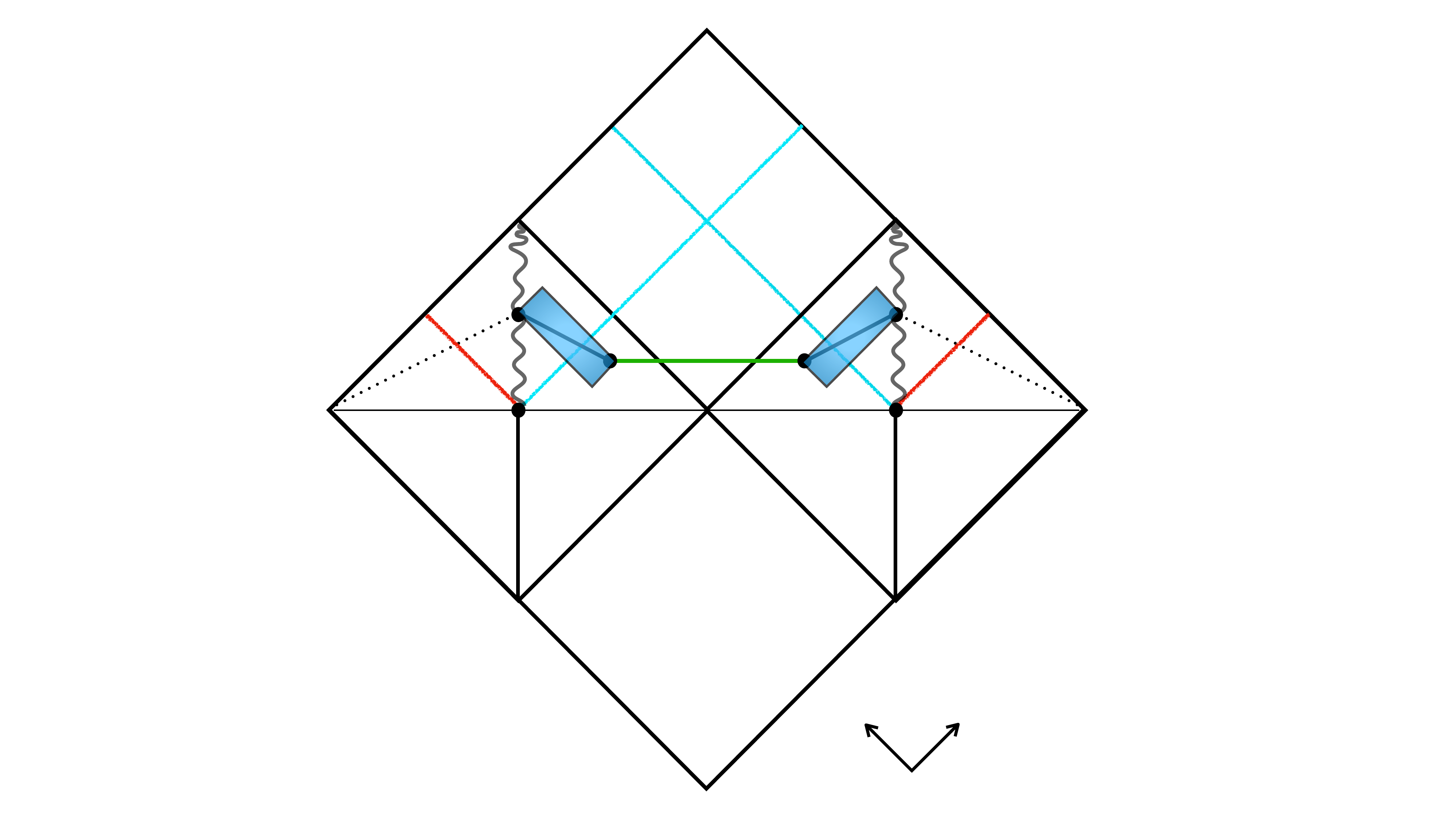}
    \put(-165,10){$w^\pm$}
    %\put(-140,102){$y_R^\pm$}
    \put(-380,125){$-\infty_{\rm L}$}
    \put(-110,125){$\infty_{\rm R}$}
    \put(-247,160){{\rm Island}}    
    \put(-309,115){$\phi_L$}
      \put(-170,115){$\phi_R$}
       \put(-150,115){\line(3,-2){2}}
     \caption{We compute the change in entanglement entropy associated to the blue diamonds.}
\label{Fig:IslandFirstlaw}
\end{figure} 
  
We begin by analyzing the case where the quantum extremal surface $\partial I$ is outside the future light cone of the shock wave, as illustrated in Fig. \ref{Fig:IslandFirstlaw}. We then compute the generalized entropy for this configuration and look for non-trivial saddles. 

Upon examining the dilaton solution for an excited state (\ref{eq:phigeneral}), we observe that the dilaton profile remains unaltered after backreaction
\be
\delta \phi = 0 ~.
\ee
Consequently, the area term in the genralized entropy is unaffected. Thus, the shockwave contribution is entirely encoded in the matter entropy term $S_{\rm matt}(R\cup I)$. To compute the change in the matter entropy, we apply the first law of entanglement to the complementary intervals $(R\cup I)^c$, or equivalently the diamond regions depicted in Figure \ref{Fig:IslandFirstlaw}). The variation in the matter entropy is thus given by the formula 
\be\label{eq:firstlaw}
\delta S_{\rm matt} = \int d \Sigma_\mu \delta \langle T^{\mu \nu}\rangle K_\nu~,
\ee
where the perturbed stress tensor is given by (\ref{eq:Tnew}), $K_{\mu}$ is the killing vector, which preserves the shape of the causal diamond, and $d\Sigma_m$ is the proper volume element on $\Sigma$. 

The non-local coupling excites the initial vacuum state of the field theory
\be
\ket{\omega} \rightarrow  \ket{\omega} + \delta \ket{\Psi}~.
\ee
Since the original stress tensor corresponds to the global vacuum state ($T_{w^\pm w^\pm}=0$), the perturbed stress tensor components are determined by the expression computed in the previous section (\ref{eq:Tnew}). Importantly, the Weyl anomaly is present both in the original and final stress tensors, ensuring its cancellation and yielding a finite stress tensor fluctuation in (\ref{eq:firstlaw}). 

The conformal Killing vector preserving the diamond region with future and past tips at $(y^\mu, x^\mu)$ takes the following form \cite{deBoer:2016pqk}
\be
\label{eq:conformal}
K^\mu(w)\partial_\mu = -\frac{ 2\pi}{(y-x)^2} \left[(y-w)^2 (x^\mu - w^\mu) - (x-w)^2(y^\mu-w^\mu) \right]\partial_\mu~. 
\ee
This vector vanishes at $w^\mu =x^\mu$ and $w^\mu = y^\mu$, as well as when $(y-w)^2=0$ or $(x-w)^2=0$. In two dimensions, it is convenient to to express it in light-cone coordinates
\be
u = t+x ~, ~ v = t - x ~,
\ee
to simplify its components 
\be
\label{eq:killing2D}
K^u(u) = 2\pi \left( \frac{(u-u_r)(u-u_I)}{u_I-u_r} \right)~, ~~ K^v(v) = 2\pi \left( \frac{(v-v_I)(v_r-v)}{v_I-v_r} \right)~,
\ee
where the left and right endpoints of the diamond have coordinates $(u_I,v_I)$ and $(u_r, v_r)$, respectively. As a consistency check, we can take the Rindler limit in which the endpoints of the diamond become $(u_I,v_I)=(0,0)$ and $(u_r,v_r)=(\infty, -\infty)$. The conformal Killing vector components become
\be
K^u \approx u ~, {\rm and} ~~ K^v \approx -v ~,
\ee
reproducing the known Minkowski result.
\be
K = t \partial_x + x \partial_t = u \partial_u - v \partial_v~.
\ee
Going back to the matter entropy computation, we can use the normal vector to the tilted space-like interval $(R \cup I)^c$
\be
n^\mu = \left(\sqrt{\frac{u_r-u_I}{v_I-v_r}},\sqrt{\frac{v_I-v_r}{u_r-u_I}}\right)~,
\ee
and evaluate
(\ref{eq:firstlaw}) for the diamond in the right black hole exterior using (\ref{eq:killing2D}) and (\ref{eq:Tnew}). Thus, we obtain the expression
\be
\delta S_{\rm matt}((R \cup I)^c) = -\frac{ g }{2} \left[\frac{(u_R -u_I)(u_R-u_r)}{(u_R - u_L)(u_I-u_r)} \right]~.
\ee
Similarly, we can compute (\ref{eq:firstlaw}) for the left black hole exterior 
\be
\delta S_{\rm matt}((R \cup I)^c) = -\frac{g}{2} \left[\frac{(v_L -v_I')(v_L-v_r')}{(v_L - v_R)(v_I-v_r')} \right]~.
\ee
Importantly, both matter entropies are finite as expected. 

Putting all the terms together, we obtain the generalized entropy 
\be
\label{eq:Sgenshock1}
S_{\rm gen} = \frac{\phi(w_1^+,w_1^-)}{2 G_N} + \frac{c}{3} \ln\left(\frac{- \Delta w^+ \Delta w^-}{\epsilon_1 \epsilon_2 \Omega(w_1^+,w_1^-) \Omega(w_2^+, w_2^-)}\right) + 2 ~ \delta S(w_1^+,w_1^-)~.
\ee
Extremizing with respect to the QES location $(w^+_I, w^-_I)$, we can find the location of an island to first order in the non-local coupling $g$ and in the semiclassical limit $G_N \ll 1$
\be
\label{eq:QES1}
\omega^{+}_I=-\frac{c\ell^2 \beta G_N}{24\pi \phi_r \omega_r^{-}} + \frac{c\ell^2 \beta G_N^2 g}{96\pi \phi_r} \frac{(w^+_r - w^+_R)^2}{w^+_L-w^+_R}~,~~~\omega^{-}_I=-\frac{c\ell^2 \beta G_N}{24\pi \phi_r \omega_r^{+}}- \frac{c\ell^2 \beta G_N^2 g}{96\pi \phi_r} \frac{(w^+_r - w^+_R)^2}{w^+_L-w^+_R}~.
\ee
In the limit $g \rightarrow0$, we recover the result for the global vacuum (\ref{eq:QESsaddle}) $\ket{\omega}$. Evaluating the saddle (\ref{eq:QES1}) in the the generalized entropy (\ref{eq:Sgenshock1}), we find
\be \label{eq:reductionEntropy}
S_{\rm gen} = 2S_{BH} -  \frac{g (w_r^+ - w^+_R)w^+_R}{w^+_r(w^+_L-w^+_R)}~.
\ee
Therefore, the entropy of Hawking radiation decreases due to the non-local coupling, consistent with the fact that the black hole horizon shrinks as negative energy falls into it. 
Since in this configuration the entropy of the Hawking radiation lies in the flat region of the Page curve, the result can also be interpreted as the point where the radiation entropy matches the total entropy of the black hole, which itself has been reduced by the negative energy flux. This is directly analogous to the reduction of black hole entropy in the GJW setup, equation~\eqref{eq:deltaS-GJW}, obtained there for a massive scalar field in AdS$_2$.  As time evolves, the QES eventually reaches the shockwave insertion point at $(w^+_s,w^-_s)$ and enters its future light cone. We analyze this scenario in detail in the following section.

\subsubsection{$\partial I$ inside the shock wave future light cone}
\label{sec:inside}

We now investigate the situation in which the candidate QES enters the future light cone of the non-local coupling insertion. In this configuration, the shockwave does not cross the complementary interval $(R \cup I)^c$ and remains outside its causal diamond. As a result, the matter entropy remains unaffected by the shock wave
\be
\delta S_{\rm matt} = 0 ~.
\ee
The entire change in generalized entropy is therefore determined by the modification of the area term through the dilaton solution (\ref{eq:deltaphi}) gets modified. 

For computational convenience, we express the dilaton in mixed coordinates $(y^+,w^-)$, which explicitly incorporate the gluing map $x(y^+)$ \cite{Goto:2020wnk, Hollowood:2020cou}. Moreover, we focus on the right exterior of the black hole where the dilaton then becomes
\be
\phi(y^+,w^-) = \phi_0 + 2 \phi_r \left(\frac{x''(y^+)}{2 x'(y^+)} - \frac{w^- x'(y^+)}{1+w^- x(y^+)} \right)~, ~ y^+ := x^{-1}(x^+)~.
\ee
Here, we have extended the gluing map into the gravitating region by matching the coordinates in the rigid bath region to the exterior black hole across the AdS boundary
\be
w^{\pm} = \pm \ell e^{\pm 2\pi y_R^{\pm}/\beta}~, ~~ w^{\pm} = \mp \ell e^{\mp 2\pi y_L^{\pm}/\beta}~.
\ee
After the shockwave backreaction, we assume that the right Hawking modes are unaffected, \ie they are still in thermal equilibrium, whereas the left moving modes change according to the gluing map. Thus,
\be\label{eq:map1}
w^+ =\ell e^{\frac{2\pi y^+}{\beta}}~, ~~~ w^- = -\ell/x(y^-)~.
\ee
The conformal factors involved in the generalized entropy computation will be also modified. They have the form 
\be
\Omega^2(w_1^-,y^+_1)^2 = \frac{\pi}{2 \beta}\frac{e^{\frac{\pi y_1^+}{\beta}}}{x'(y_1^+)} (1+ x(y_1^+) w^-_1)^2 ~, ~~ \Omega^2 (w_2^-,y^+_2) = \frac{2\pi}{\beta}\frac{x'(t)}{x(t)^2} e^{\frac{2\pi t}{\beta}}~.
\ee
We can then evaluate the generalized entropy
\be\label{eq:Sgendilaton}
S_{\rm gen} = \frac{\phi_0}{4 G_N} + \frac{\phi_r}{4 G_N} \left(\frac{x''(y_1^+)}{2 x'(y_1^+)} - \frac{w^-_1 x'(y_1^+)}{1+w_1^- x(y_1^+)} \right) + \frac{c}{6} \ln \left( \frac{- \Delta w^+ \Delta w^-}{\epsilon_1 \epsilon_2 \Omega_1(w_1^-,y^+_1) \Omega_2(w_2^-,y^+_2)} \right) ~.
\ee
The extremization with respect to $w^-$ can be easily solved
\be
\label{eq:w1saddle}
\partial_{w_1^-} S_{\rm gen} = 0 ~ \implies ~ w_1^- = \frac{x'(y^+_1) + k x(y_1^+) - k x(t)}{x(t) (k x(y_1^+) - x'(y_1^+)) - kx(y_1^+)^2}
\ee
At late times $t \gg t_0$, we can use the asymptotic expansion for the gluing map (\ref{eq:xlate}), and approximate the value of $w^-_1$ in (\ref{eq:w1saddle}). Furthermore, we can solve in this regime for the QES location in the semiclassical limit $k \ll 1$. This procedure results in the QES location at
\be\label{eq:qeslate}
w_1^- \approx 
\frac{g}{4 e^{\frac{2\pi t_0}{\beta}}(1-2 e^{\frac{2\pi }{\beta}(t_0-t)})}~, ~~ y_1^+ \approx t + \frac{\beta}{4 \pi} \ln \left(\left(\frac{\pi}{\beta k }\right)^2 + 1 \right) ~.
\ee
From these expressions we can read off the scrambling time
\be
\Delta t_s = \frac{\beta}{4 \pi} \ln \left(\left(\frac{\pi}{\beta k }\right)^2 + 1 \right)~,
\ee
which is bigger that the original scrambling time in the unperturbed state $\ket{\omega}$. It is worth mentioning that this result is consistent with the idea that the island shrinks a little bit inward from the original horizon. In the extremal limit, the scrambling time is $\Delta t_s = \pi / 4 \beta k^2$, while at high temperatures $\Delta t_s = \frac{\beta}{2\pi} \ln \left(\frac{\pi}{\beta k} \right)$. We see that at late times the island is finite distance away from the horizon and approaches the point $w^+ \rightarrow \infty$ as $t \rightarrow \infty$. Finally, using (\ref{eq:qeslate}) we evaluate the generalized entropy
\be\label{eq:Slate}
S_{\rm gen} = \frac{\phi_0}{2 G_N} - \frac{\phi_r}{2 G_N}\frac{\pi}{\beta} + \frac{2 c}{3} \ln \frac{\beta}{\pi} + \dots ~.
\ee
We see that the black hole entropy decreases with respect to the original black hole entropy. Using the late time approximmation for the gluing map (\ref{eq:xlate}), we can compute Bekenstein-Hawking entropy for the black hole after backreaction at late times
\begin{align}
S_{BH}= \frac{\phi(y^+,w^-)}{4 G_N} \Big\lvert_{\rm hzn} & = \frac{\phi_0}{4 G_N} + \frac{\phi_r}{4 G_N} \frac{x''(y^+)}{x'(y^+)} \\ \nonumber
& \approx \frac{\phi_0}{4 G_N} - \frac{c \nu}{12} -\frac{c u_0^2}{12}(1+\nu) e^{-k(y^+-L)} \\ \nonumber
&=  \frac{\phi_0}{4 G_N} - \frac{\phi_r}{4 G_N}\frac{2 \pi}{\beta} -  \frac{c \pi}{3 \beta k } + \frac{c}{3} \ln \frac{\beta}{\pi} - \frac{c \nu u_0^2}{12}~.
\end{align}
This entropy constitutes a non-trivial check of the result (\ref{eq:Slate}). We thus have shown that even after backreaction, there is a non-trivial island, which at late times recovers the Page curve.

\section{Correlators in the dual quantum mechanical model}
\label{sec5}
The gravitational system considered in this work, where matter fields are coupled to gravity, admits a dual quantum mechanical description at the boundary separating the gravitating region from the thermal baths \cite{Almheiri:2019hni}. In this section, we investigate the impact of negative energy shock waves on two-point correlation functions within this dual description. The operators in these correlation functions correspond to matter fields in the gravitational theory, and we compute these correlators by working on the gravity side and applying the AdS/CFT dictionary.

Free matter fields in the gravitating region with boundary conditions $\chi_b(t)$ can be shown to have the following effective action~\cite{Maldacena:2016upp}
\begin{equation}
    I_\text{eff}=-c_\Delta \int dt \,dt' \left( \frac{F'(t)F'(t')}{\left[ F(t)-F(t')\right]^2} \right)^\Delta \chi_b(t) \chi_b(t')\,,
\end{equation}
where $\Delta$ is the scaling dimension of the dual boundary operator $\mathcal{O}$, and $c_\Delta=\frac{(\Delta-1/2)\Gamma(\Delta)}{\sqrt{\pi}\,\Gamma(\Delta-1)}$.

The two-point function between two boundary operators can be shown to be given by~\cite{Maldacena:2016upp}
\begin{equation} \label{eq-formulaTwoPointFunction}
    G_2(t,t')=\langle \mathcal{O}(t)\mathcal{O}(t') \rangle= c_\Delta \left( \frac{F'(t)F'(t')}{\left[ F(t)-F(t')\right]^2} \right)^\Delta
\end{equation}
where the gluing map $F(t)$ solves the equations of motion derived from the Schwarzian action $I_\text{Sch}=- C \int dt \, \{F,t \}$ with the appropriate boundary conditions (\ref{eq:JTbdycond.}).

Before the shocks, it is convenient to parametrize the the gluing map as follows:
\begin{equation}
    F(t)=\frac{\beta}{\pi} \tanh \left( \frac{\pi}{\beta}t \right)\,, ~~~~ t\leq t_0\,,
\end{equation}
which leads to the following two-point function
\begin{equation}
    G_2(t,t')=\left( \frac{\pi}{\beta} \right)^{2\Delta} \frac{c_{\Delta}}{\left(\sinh \left[ \frac{\pi}{\beta}(t-t') \right] \right)^{2\Delta}}\,,~~~~ t\leq t_0\,.
\end{equation}
For practical convenience, we work with a normalized, symmetrical two-sided correlator, which can be obtained using the expression above by setting $t' \rightarrow -t +i\, \beta/2$:
\begin{equation} \label{eq-nomalized2pfunction}
    g_2(t):=\frac{G_2(t,-t+i \, \beta/2)}{G_2(0,i \, \beta/2)}={\left(\cosh \left[ \frac{2\pi}{\beta}t \right] \right)^{-2\Delta}}\,,~~~~ t\leq t_0\,.
\end{equation}

Once the shock waves are introduced, the gluing map changes as $F(t)\rightarrow F_s(t)$, with~\cite{Hollowood:2020cou}
\begin{equation} \label{eq-NewGluingMap}
    F_s(t)=\frac{\beta}{\pi}\frac{K_{\nu}(\nu z_0) \left(f(t)/f(t_0)-1 \right)+z_0 \tanh \frac{\pi t_0}{\beta}\left(f(t)I_{\nu}'(\nu z_0)/\alpha-K_{\nu}'(\nu z_0) \right)}{K_{\nu}(\nu z_0) \left(f(t)/f(t_0)-1\right)\tanh \frac{\pi t_0}{\beta}+z_0\left(f(t)I_{\nu}'(\nu z_0)/\alpha-K_{\nu}'(\nu z_0) \right)}\,,
\end{equation}
where
\begin{equation}
    f(t) = \alpha \frac{K_{\nu}(\nu z)}{I_{\nu}(\nu z)}\,,~~~~z=\sqrt{\frac{E_s}{E_\beta}}e^{-k(t-t_0)/2}\,.
\end{equation}
Here, $\alpha$ is a normalization constant, which we fix to $\alpha = 1$. The new gluing map \eqref{eq-NewGluingMap} was originally derived in~\cite{Hollowood:2020cou} for positive-energy shock waves. Interestingly, we find that this expression also yields consistent results for negative-energy shock waves, which can be obtained via the analytic continuation $z \rightarrow i z$. This continuation arises naturally when $E_s < 0$ is substituted under the square root. 

\begin{figure}
    \centering
    \includegraphics[width=0.7\linewidth]{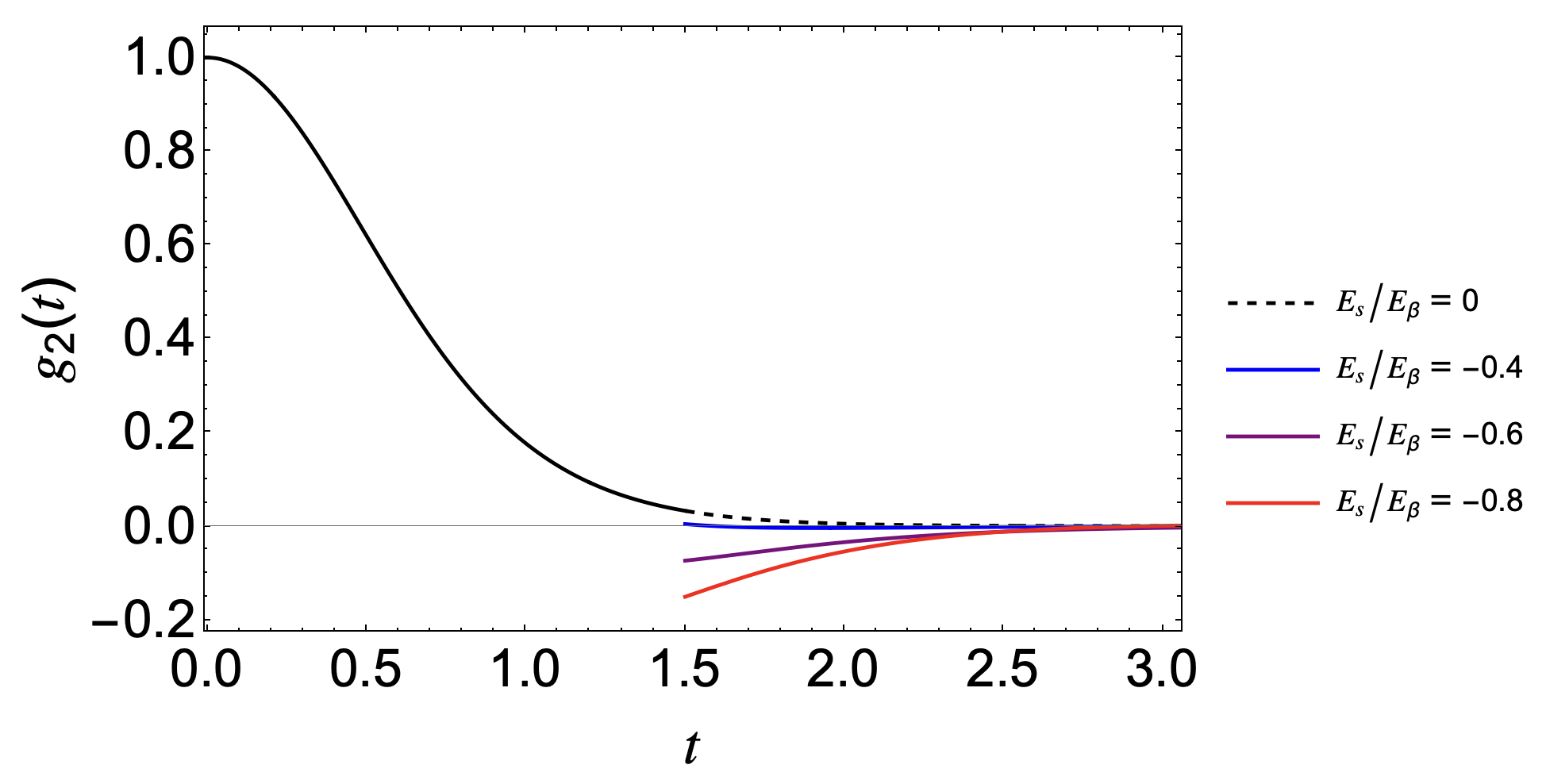}
    \caption{Impact of negative-energy shock waves on the normalized two-sided two-point function $g_2(t)$, as defined in Eq.~(\ref{eq-nomalized2pfunction}). Shock waves with varying energy, distinguished by color, are introduced at $t_0 = 1.5$, resulting in a discontinuity whose magnitude increases with increasing $|E_s|$. At later times $t \gg t_0$, the perturbed two-point functions converge back to the unperturbed result (dashed line), following the same trajectory regardless of the shock wave energy.
}
    \label{fig:TwoPointFunction}
\end{figure}

By substituting \eqref{eq-NewGluingMap} into \eqref{eq-formulaTwoPointFunction}, we obtain the two-point function of matter fields after the shock. Figure~\ref{fig:TwoPointFunction} illustrates the effect of negative-energy shock waves on the normalized two-sided two-point function $g_s(t)$. The insertion of shock waves at \( t = t_0 \) induces a discontinuity in the two-point function, whose magnitude grows with \( |E_s| \); however, the perturbed two-point function gradually converges back to the unperturbed result at later times.

It is natural to interpret the two-sided correlator $g_s(t)$ as a direct measure of correlations between the left and right thermal baths, which in the gravitational picture correspond to the outgoing Hawking radiation on both sides. The discontinuity induced by the shock reflects how the non-local coupling modifies these correlations, while the subsequent relaxation shows that the baths eventually reestablish their unperturbed entanglement structure. In the context of recently proposed teleportation protocols realized in entangled SYK-like systems, one may view the presence of the baths as modeling environmental effects, under the assumption that the SYK setups are not fully isolated. In such scenarios, we expect the baths would influence correlations in a qualitatively similar manner to the effect analyzed in this section.

\section{Discussion}

In this work, we revisited a model of black hole evaporation exhibiting a duality between the radiation system, comprising quantum fluctuations in the non-gravitating region $\mathcal{R}$, and entanglement islands in the gravitating region encoding the black hole interior, \ie the island/radiation duality. Our primary contribution is an operational protocol for information recovery from the island through non-local coupling of simple operators in the entanglement wedge of radiation $W[\cal{R}]$. This coupling generates shockwaves that backreact on the gravitating system, effectively opening a wormhole and enabling information to escape. Crucially, our protocol does not rely on holography, as solely depends on radiation field couplings that excite the quantum state of the Hawking radiation.
  
We further investigated how negative energy modifies the generalized entropy of Hawking radiation in the gravitating theory. Our analysis reveals that there is a non-trivial island saddle after backreaction, which restores the Page curve at late times preserving unitarity. 

From the dual quantum mechanical viewpoint, we analyze correlation functions of boundary fields $g\,\varphi_{\bf L}\varphi_{\bf R}$, initially in a thermal state. Their evolution under deformations of the gluing map captures the impact of the non-local coupling and provides a direct probe of information transfer from the island to the exterior. The two-sided correlator $g_s(t)$ thus quantifies correlations between the left and right thermal baths, corresponding in the gravitational picture to Hawking radiation on both sides. The discontinuity created by the shock encodes the immediate effect of the coupling, while the subsequent relaxation indicates that the baths gradually restore their entanglement structure. In the context of SYK-like quantum simulators implementing teleportation protocols, the inclusion of auxiliary baths could effectively model environmental influences, with qualitative effects on correlations similar to those described here.

The scope of our protocol may extend to other observables when coupled to an auxiliary reservoir. For instance, higher-point correlation functions \cite{Haehl:2021tft} could broaden its applicability by probing further into the island interior geometry. The inclusion of a reservoir also makes such observables more physically realistic, as the bath can model environmental effects, thereby opening the door to further explorations, for example in the study of chaos via OTOCs.

The JT/SYK system has been studied in the lab, e.g. the sparse SYK model experimentally realized by Google Sycamore~\cite{Jafferis:2022crx}. In principle our protocol can be realized in the lab, as the state of the wormhole corresponds to the ground state of simple Hamiltonian. 

The island/radiation duality remains to be fully understood. In principle, reconstruction methods learned in AdS/CFT, such as HKLL reconstruction formulas should provide information about local operators in the island region. A key question is how these methods are modified in the presence of a reservoir. Our results suggest the HKLL formula with fluctuating boundary would be modified as
\be
\label{eq:HKLL}
\varphi_I(z,t) = \int dt' K \Big( z, F[t]~ \lvert~ F[t'] \Big) \mathcal{O}_{R}(F[t'])~,
\ee
where $K$ is the bulk to boundary propagator, $F[t]$ accounts for the boundary dynamics induced by the reservoir coupling and the local operator belongs to the entanglement wedge of radiation $\mathcal{O}\in W[{\cal{R}}]$.

Our work makes apparent a property of the fields in the radiation region: acting with simple operators in $W[\cal{R}]$ generates backreaction that effectively ``creates'' spacetime; in our case a wormhole. This raises important questions about entanglement wedge reconstruction of the island region in more general situations, This represents an important challenge, specially 
since such reconstruction must handle exponentially growing complexity. We hope to address these challenges in future work.

\acknowledgments
It is a pleasure to thank Paul Balavoine, Ben Freivogel and Dora Nikolakopoulou for discussions in the early stages of this project. We thank César A. Agón, Luis Apolo, Horacio Casini, Bart{\l}omiej Czech, Kanato Goto, Jeremy van der Heijden, Wen-Xin Lai, Raghu Mahajan, Juan F. Pedraza, Debajyoti Sarkar, Amirhossein Tajdini, and Zhenbin Yang for useful discussions. RE thanks the Gwangju Institute of Science and Technology (GIST) and the Shanghai Institute for Mathematics
and Interdisciplinary Sciences (SIMIS) for hospitality while this work was completed. R.~Esp\'indola is supported by the Dushi Zhuanxiang Fellowship and acknowledges a Shuimu Scholarship as part of the Shuimu Tsinghua Scholar Program. 
V.~Jahnke was supported by the Basic Science Research Program through the National Research Foundation of Korea (NRF), funded by the Ministry of Education under grant RS-2023-00248186, and by the Conselho Nacional de Desenvolvimento Científico e Tecnológico (CNPq), Brazil, under grant Processo 446326/2024-0 (Bolsa Conhecimento Brasil – BCB-1). 
This work was supported by the Basic Science Research Program through the National Research Foundation of Korea (NRF) funded by the Ministry of Science, ICT \& Future Planning (NRF-2021R1A2C1006791), the framework of international cooperation program managed by the NRF of Korea (RS-2025-02307394), the Creation of the Quantum Information Science R\&D Ecosystem (Grant No. RS-2023-NR068116) through the National Research Foundation of Korea (NRF) funded by the Korean government (Ministry of Science and ICT), the Gwangju Institute of Science and Technology (GIST) research fund (Future leading Specialized Resarch Project, 2025) and the Al-based GIST Research Scientist Project grant funded by the GIST in 2025. This research was also supported by the Regional Innovation System \& Education(RISE) program through the (Gwangju RISE Center), funded by the Ministry of Education(MOE) and the (Gwangju Metropolitan City), Republic of Korea.(2025-RISE-05-001)

\appendix

\section{Evaporating 2D black hole}
\label{app:A}
Following the conventions of \cite{Goto:2020wnk}, by a similar procedure as in the main text we can rewrite the Schwarzian equation of motion as the modified Bessel equation
\be
\left(y^2 \frac{d^2}{dy^2} + y \frac{d}{dy}  - (y^2 + \nu^2)\right) F =0~, ~~ \nu = \frac{2\pi}{k \beta}~~~,
\ee
where
\be
F := e^{- \frac{1}{2}\phi}~,~~~ y := \nu \sqrt{\frac{E_\psi}{E_{BH}}} e^{-\frac{k}{2}(t-t_0)}~,~~~ {\rm and} ~~\phi := \log x'(t)~.
\ee
The solution is then in terms of modified Bessel functions of the first and the second kinds respectively
\be
F(y) = c_1 K_\nu(y) + c_2 I_\nu (y)~.
\ee
Inverting we get, 
\be\label{eq:sol}
x(t) =  \int\limits_{t_0}^t dt' \frac{1}{\left[c_1 K_\nu(y(t')) + c_2 I_\nu (y(t'))\right]^2} +c_3 ~.
\ee
Imposing the matching conditions
\be
x(t_0) = e^{\frac{2\pi}{\beta}t_0}~,~~ x'(t_0) =\frac{2\pi}{\beta} e^{\frac{2\pi}{\beta}t_0} , ~~ {\rm and}~~ x''(t_0)  = \left(\frac{2\pi}{\beta}\right)^2 e^{\frac{2\pi}{\beta}t_0} ~,
\ee
we fix the constants
\begin{align}
c_1 &=  \frac{e^{-\frac{\pi }{\beta}t_0}}{k \sqrt{2\pi\beta}} \left[(2 \pi + k \beta \nu) I_{\nu}(y_0) - ky_0 \beta I_{\nu-1}(y_0)\right]~, \\ \nonumber
c_2 &= -  \frac{e^{-\frac{\pi }{\beta}L}}{k \sqrt{2\pi\beta}} \left[(2 \pi + k \beta \nu)  K_\nu(y_0) + ky_0 \beta K_{\nu-1}(y_0) \right]~, \\ \nonumber
c_3 &= e^{\frac{2\pi}{\beta}t_0}~,
\end{align}
where we defined $y(t_0) :=y_0$. 

In \cite{Goto:2020wnk}, the solution for $t>t_0$ is parametrized as follows
\be\label{eq:solGHT}
X(t) =e^{\frac{2\pi }{\beta}t_0} \left[1 + \frac{2}{u_0} \frac{-K_\nu(\nu u_0) I_\nu(\nu u) + I_\nu(\nu u_0) K_\nu(\nu u)}{K_{\nu+1} (\nu u_0) I_\nu(\nu u) + I_{\nu+1} (\nu u_0) K_\nu (\nu u)} \right] ~,
\ee
where $u = u_0 e^{- \frac{k}{2}(t-t_0)}$, $u_0 = \beta \sqrt{\frac{12 k E_\psi}{c\pi}}$ and $\nu = \frac{2\pi}{\beta k}$.

\begin{figure}
\begin{center}
\includegraphics[width=0.6\textwidth]{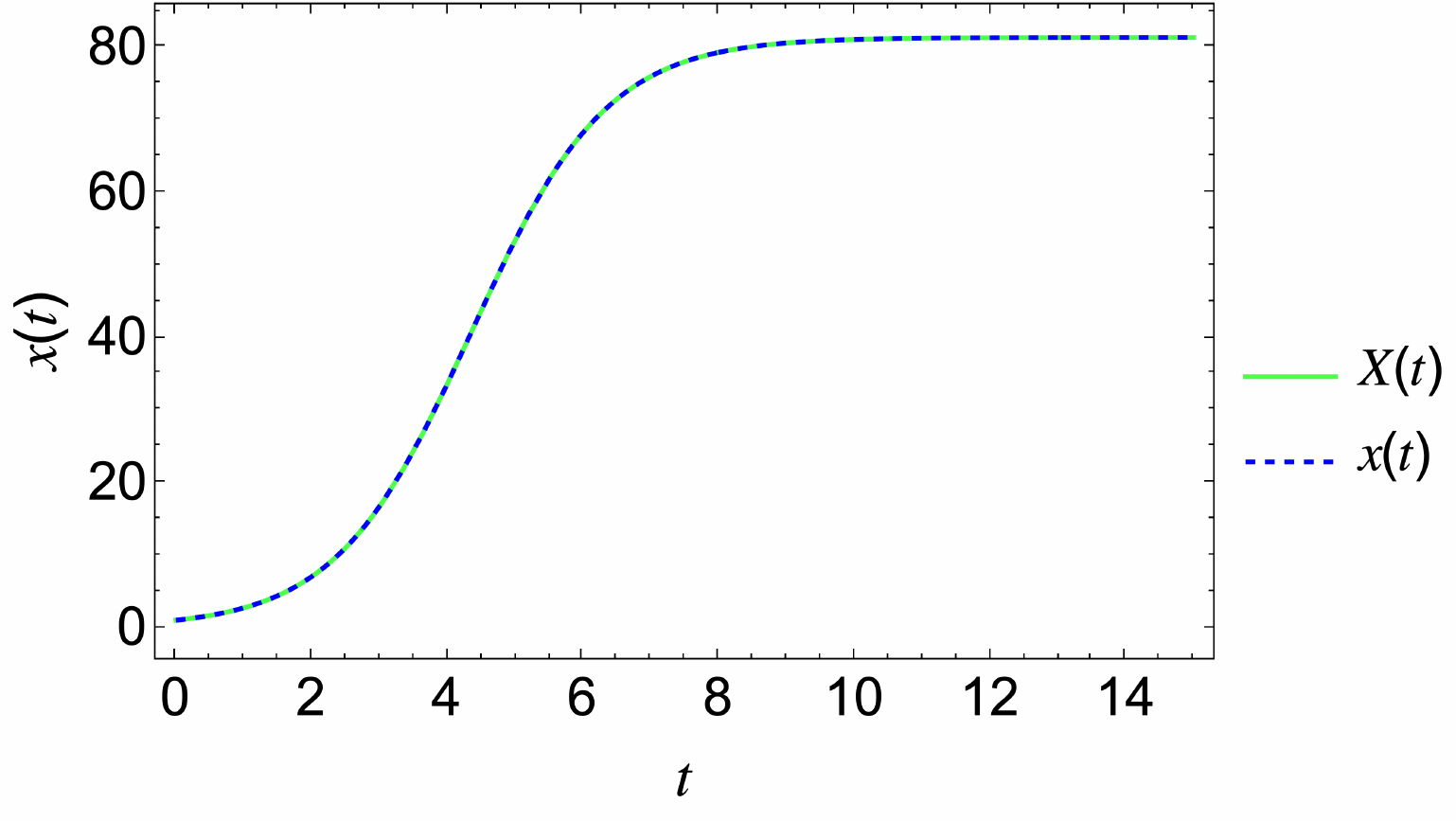}
\caption{Gluing map solutions  (\ref{eq:sol}) and (\ref{eq:solGHT}) with parameters $L=0$, $u_0=1$, $\nu=1$, $\beta = 2 \pi$ and $k=1$.}
\label{Fig:Bessels}
\end{center}
\end{figure}

In fact, solutions (\ref{eq:sol}) and (\ref{eq:solGHT})  are the same, as the relation between parameters  is given by
\be
y = \nu u~, ~~ u_0 = \sqrt{\frac{E_\psi}{E_{BH}}}~. 
\ee
This can be seen numerically in Figure \ref{Fig:Bessels}.

\section{Diagnosing traversability with two-sided commutator}
\label{app:B}
In this appendix, following \cite{Maldacena:2017axo}, we review how the traversability of Gao–Jafferis–Wall wormholes can be diagnosed through two-sided commutators, restricting for simplicity to an AdS$_2$ background.

We open the wormhole, considering a double-trace deformation of the form
\begin{equation}
    \mathcal{V}=\frac{1}{K}\sum_{j=1}^{K}\mathcal{O}^j_L(-t_0) \mathcal{O}^j_R(t_0)\,,
\end{equation}
where all the $K$ fields have the same scaling dimension, $\Delta$. By a suitable choice of the sign of the coupling $g$, the boundary operators generate negative energy shock waves in the bulk. 

\begin{figure}[]
\centering
%\captionsetup{justification=centering}

\begin{tikzpicture}[scale=1.5]
\draw [thick]  (0,0) -- (0,3);
\draw [thick]  (3,0) -- (3,3);
\draw [thick,dashed]  (0,0) -- (3,3);
\draw [thick,dashed]  (0,3) -- (3,0);

\draw [thick,blue,decorate,decoration={snake,segment length=3mm,amplitude=0.5mm}]  (3,1.5) -- (1.5,2.9);
\draw [thick,blue,decorate,decoration={snake,segment length=3mm,amplitude=0.5mm}]  (3.03,1.53) -- (1.53,2.93);

\draw [thick,blue,decorate,decoration={snake,segment length=3mm,amplitude=0.5mm}]  (0,1.5) -- (1.5,2.9);
\draw [thick,blue,decorate,decoration={snake,segment length=3mm,amplitude=0.5mm}]  (-0.03,1.53) -- (1.47,2.93);

\draw [red,thick,->]  (0,0.2) -- (1.54,1.74);
\draw [red,thick]  (1.54,1.74) -- (2.14,2.34);
\draw [red,thick,->]  (2.34,2.14) -- (2.7,2.5);
\draw [red,thick]  (2.7,2.5) -- (3,2.8);

\draw [thick,decorate,decoration={zigzag,segment length=1.5mm, amplitude=0.3mm}] (0,3) .. controls (.75,2.85) 
and (2.25,2.85) .. (3,3);
\draw [thick,decorate,decoration={zigzag,segment length=1.5mm,amplitude=.3mm}]  (0,0) .. controls (.75,.15) and (2.25,.15) .. (3,0);

\draw[thick,<->] (-0.4,3.4) -- (-0.1,3.1) -- (0.2,3.4);

%\node[scale=0.8, align=center] at (1.5,2.65) {Future Interior};
%\node[scale=0.8,align=center] at (1.5,.55) {Past Interior};
%\node[scale=0.8,align=center] at (0.6,1.6) {Left\\ Exterior};
%\node[scale=0.8,align=center] at (2.4,1.6) {Right\\ Exterior};
\end{tikzpicture}
%\put(10,60){\Large $= |TFD \rangle$}
\vspace{0.1cm}
\put(-150,60){\rotatebox{0}{\small $\mathcal{O}_L$}}
\put(-1,60){\rotatebox{0}{\small $\mathcal{O}_R$}}
\put(-170,10){\rotatebox{0}{\small $\phi_L(t_L)$}}
\put(-1,120){\rotatebox{0}{\small $\phi_R(t_R)$}}
\put(-157,150){\rotatebox{0}{\small $X^-$}}
\put(-125,150){\rotatebox{0}{\small $X^+$}}

\caption{\small The non-local coupling between $\mathcal{O}_L$ and $\mathcal{O}_R$ introduces a negative-energy shock wave in the bulk that makes the wormhole traversable. The traversability can be diagnosed by a two-sided correlation function (\ref{eq:Ctrav}) involving $\psi_L$ and $\psi_R$.}
\label{fig:info-tranfer}
\end{figure}
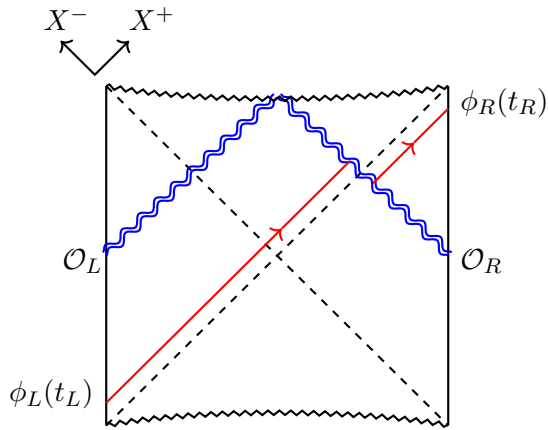

We diagnose traversability with a two-sided boundary commutator of the form
\begin{equation} \label{eq:Ctrav}
    \mathcal{C} = \langle [\phi_L,e^{-i g \mathcal{V}}\phi_R e^{i g \,\mathcal{V}}]\rangle\,
\end{equation}
where $\phi$ is a boundary operator with scaling dimension $\Delta_{\phi}$, and represents a signal that is sent though the wormhole. The commutator can also be computed as
\begin{equation}
    \mathcal{C} = 2\,i \, \text{Im}\, C\,,\,\,\,\,C=\langle e^{-i g \mathcal{V}}\phi_R e^{i g \,\mathcal{V}} \phi_L \rangle\,.
\end{equation}
Taking the limit where $K$ is very large, and using a small $G_N$ approximation, one can write
\begin{equation}
    C=e^{-ig\langle \mathcal{V} \rangle} \tilde{C}\,,\,\,\,\, \tilde{C}=\langle \phi_R \, e^{i g \,\mathcal{V}} \phi_L \rangle\, .
\end{equation}
The correlator $\tilde{C}$ can be thought of as a scattering amplitude between the bulk fields generated by the signal $\phi$ and the negative energy shocks generated by $\mathcal{O}_{L,R}$\footnote{For simplicity, we denote $\mathcal{O}^j$ by $\mathcal{O}$.}. If the wave function of the state created by $\phi$ and $\mathcal{O}$ have a large relative boost, one can model their interaction using the gravity eikonal approximation. In this case, $\tilde{C}$ can be written as follows~\cite{Maldacena:2017axo}:
\begin{equation} \label{eq-Ctilde}
    \tilde{C}=\int_0^{\infty} dp^+ \psi_\text{signal}(p^+;t_L,t_R) e^{i g \int_{0}^{\infty}dq^{-} \psi_\text{shock}(q^{-};t_0)e^{i \delta(p^+,q^-)}}
\end{equation}
where the phase shift 
\begin{equation}
    \delta(p^+,q^-) = G_N p_+ p_-\,,
\end{equation}
controls the interaction between the signal and the shock. Here we absorbed possible numerical constants into $G_N$. The wave functions are given by
\begin{align} \label{eq-wavefunctions}
    \psi_\text{signal}(p^+;t_L,t_R) &= \int_{-\infty}^{\infty} \frac{da}{2\pi}\, e^{-i ap^+} \langle \phi_R(t_R) e^{ i a P^+} \phi_L(t_L) \rangle\,,\\
    \psi_\text{shock}(q^-;t_0) &= \int_{-\infty}^{\infty} \frac{da}{2\pi}\, e^{-i a q^-} \langle \mathcal{O}_R(t_0) e^{ i a P^-} \mathcal{O}_L(-t_0) \rangle\,,
\end{align}
where $P^{\pm}$ are generators of SL(2,$\mathbb{R}$), which is the group of isometries of AdS$_2$. To precisely describe the action of these generators, it is convenient to think about AdS$_2$ in terms of embedding coordinates $(X^{-1}, X^0, X^1)$ satisfying the constraint $(X^{-1})^2+(X_0)^2-(X_1)^2=\ell^2$ in a space with metric $ds^2=-(dX^{-1})^2-(dX^{0})^2+(dX^1)^2$. 
The generators of SL(2,$\mathbb{R}$) can be defined as~\cite{Lin:2019qwu}
\begin{equation} \label{eq-generators}
    Q^a=\frac{1}{2}\epsilon^{abc} J_{bc}\,,\,\,\,\, J_{bc}=-i X_a \frac{\partial}{\partial X^b}+i X_b \frac{\partial}{\partial X^a}\,,\,\,a=-1,0,1.
\end{equation}
The generators of null translations are defined as follows
\begin{equation}
    P^{\pm}=-P_{\mp}=\frac{Q^1\pm Q^0}{2}\,.
\end{equation}
Using (\ref{eq-generators}), we obtain
\begin{equation} \label{eq-Qdiff}
   i P^{\pm} =\pm\frac{1}{2} \left(X^{\pm} \frac{\partial}{\partial X^{-1}}+2 X_{-1} \frac{\partial}{\partial X^{\mp}} \right)\,.
\end{equation}
where $X^{\pm}=X^{0} \pm X^1$.
Using (\ref{eq-Qdiff}), one can see that the operators $e^{i a P^{\pm}}$ act on $X^a$ as follows:
\begin{align}
   & e^{i a P^+}: \,\,\,\,\,\,\,\,(X^{-1}, X^+, X^-) \longrightarrow \left( X^{-1}+\frac{a}{2}X^+, X^+, X^-+a X_{-1}-\left( \frac{a}{2}\right)^2 X^+ \right)\\
   & e^{i a P^-}: \,\,\,\,\,\,\,\,(X^{-1}, X^+, X^-) \longrightarrow \left( X^{-1}+\frac{a}{2}X^-, X^++a X_{-1}-\left( \frac{a}{2}\right)^2 X^-, X^- \right)
\end{align}
Now, let's determine how these generators act on boundary two-point functions. Boundary points can be parametrized as follows\footnote{Note that this point satisfies the condition $-X^+ X^{-}-(X^{-1})^2=0$.} $(X^{-1},X^{+},X^{-})=(-\ell,\ell e^{\frac{2\pi}{\beta}t},- \ell e^{-\frac{2\pi}{\beta}t})$. Let $\mathcal{O}$ be a scalar field with scaling dimension $\Delta$, and $P=(-\ell,\ell e^{\frac{2\pi}{\beta}t},- \ell e^{-\frac{2\pi}{\beta}t})$ and $P'=(-\ell,\ell e^{\frac{2\pi}{\beta}t'},- \ell e^{-\frac{2\pi}{\beta}t'})$ be two boundary points. Then the boundary-to-boundary correlator is given by
\begin{equation}
    \langle \mathcal{O}(P) \mathcal{O}(P') \rangle = \frac{1}{( -\frac{2 P \cdot P'}{\ell^2})^{\Delta}}=\frac{1}{(2 \sinh \frac{\pi}{\beta}(t-t'))^{2\Delta}}
\end{equation}
In a two-sided black hole geometry, we can move one of these operators to the other asymptotic boundary by simply shifting the time coordinate as follows $t' \rightarrow t'+i \beta/2$. One then obtains
\begin{equation}
    \langle \mathcal{O}_R(P) \mathcal{O}_L(P') \rangle = \frac{1}{( \frac{2 P \cdot P'}{\ell^2})^{\Delta}}=\frac{1}{(2 \cosh \frac{\pi}{\beta}(t-t'))^{2\Delta}}
\end{equation}
Now we can finally derive the formula
\begin{equation}
    \langle \mathcal{O}_R(P) e^{i a P^+} \mathcal{O}_L(P') \rangle =\frac{1}{\left( 2 \cosh \frac{\pi}{\beta}(t-t')+\frac{a}{2}e^{\frac{\pi}{\beta}(t+t')} \right)^{2\Delta}}
\end{equation}
The wave functions can then be computed as
\begin{equation}
    \psi_\text{shock}(p,t_0)= \int_{-\infty}^{\infty} \frac{da}{2\pi} \,  \frac{e^{-i a q^-}}{\left[ 2 \cosh \left( \frac{2\pi}{\beta}t_0 \right)+\frac{a}{2} \right]^{2\Delta}} =-\frac{2^{2\Delta} e^{-i \pi \Delta}}{\Gamma(2\Delta)} (q^-)^{2\Delta-1} e^{- 4 i q^{-} \cosh \left( \frac{2\pi}{\beta}t_0\right)} \theta(q^-)\,,
\end{equation}
and
\begin{align}
    \psi_\text{signal}(p^+,t_L,t_R)&= \int_{-\infty}^{\infty} \frac{da}{2\pi} \,  \frac{e^{-i a p^+}}{\left[ 2 \cosh \frac{\pi}{\beta}(t_R-t_L)+\frac{a}{2}e^{\frac{\pi}{\beta}(t_L+t_R)} \right]^{2\Delta_{\phi}}} \nonumber \\
    &=\frac{2^{2\Delta_{\phi}} e^{- \Delta \frac{2\pi}{\beta}(t_L+t_R)} e^{-i \pi \Delta_{\phi}}}{\Gamma(2\Delta_{\phi})} (p^+)^{2\Delta_{\phi}-1} e^{- 4 i p^{+} \cosh  \frac{\pi}{\beta} (t_R-t_L) e^{-\frac{\pi}{\beta}(t_R+t_L)}} \theta(p^+)\,,
\end{align}
The integral in $q^-$ in (\ref{eq-Ctilde}) can be computed as follows:
\begin{align}\label{eq-integralShock}
    \int_{0}^{\infty}dq^{-} \psi_\text{shock}(q^{-};t_0)e^{i \delta(p^+,q^-)} &=\int_{0}^{\infty}dq^{-} e^{i G_N p^+ q^-} \int_{-\infty}^{\infty} \frac{da}{2\pi}\, e^{-i a q^-} \langle \mathcal{O}_R(t_0) e^{ i a P^-} \mathcal{O}_L(-t_0) \rangle\, \nonumber \\
    &=\int_{-\infty}^{\infty} \frac{da}{2\pi} \int_{0}^{\infty}dq^{-} e^{i (G_N p^+- a)q^-}\langle \mathcal{O}_R(t_0) e^{ i a P^-} \mathcal{O}_L(-t_0) \rangle \nonumber\\
    &= \langle \mathcal{O}_R(t_0) e^{ i G_N p^+ P^-} \mathcal{O}_L(-t_0) \rangle  = \left[ 2 \cosh \left( \frac{2\pi}{\beta}t_0 \right)+\frac{G_N p^+}{2} \right]^{-2\Delta}\,.
\end{align}
Note that if we set $G_N=0$ in the integral in (\ref{eq-integralShock}), we obtain:
\begin{equation}
    \langle \mathcal{V} \rangle =\left[ 2 \cosh \left( \frac{2\pi}{\beta}t_0 \right) \right]^{-2\Delta}\,.
\end{equation}
The correlator $\tilde{C}$ can then be computed as
\begin{equation}
    \tilde{C}=  c_{\Delta_{\phi}} e^{- \Delta_{\phi} \frac{2\pi}{\beta}(t_L+t_R)} \int_0^{\infty} dp^+ \,  (p^+)^{2\Delta_{\phi}-1} e^{- 4 i p^{+} \cosh  \frac{\pi}{\beta} (t_R-t_L) e^{-\frac{\pi}{\beta}(t_R+t_L)}}  e^{i g \,\left[ 2 \cosh \left( \frac{2\pi}{\beta}t_0 \right)+\frac{G_N p^+}{2} \right]^{-2\Delta}} \,,
\end{equation}
where $c_{\Delta_{\phi}}= \frac{2^{2\Delta_{\phi}} e^{-i \pi \Delta_{\phi}}}{\Gamma(2\Delta_{\phi})}$\,. Finally, the two-sided correlator $C=e^{-ig \langle \mathcal{V}\rangle} \tilde{C}$ can be written as
\begin{equation} \label{eq-Cnumerical}
    C=  c_{\Delta_{\phi}} e^{- \Delta_{\phi} \frac{2\pi}{\beta}(t_L+t_R)}  \int_0^{\infty} dp^+ \,  (p^+)^{2\Delta_{\phi}-1} e^{- 4 i p^{+} \cosh  \frac{\pi}{\beta} (t_R-t_L) e^{-\frac{\pi}{\beta}(t_R+t_L)}}   e^{i g D(p^+) }\,.
\end{equation}
where 
\begin{equation}
    D(p^+)= \left[ 2 \cosh \left( \frac{2\pi}{\beta}t_0 \right) \right]^{-2\Delta} \left[ \left( 1+\frac{G_N p^+}{4 \cosh \left( \frac{2\pi}{\beta}t_0 \right)} \right)^{-2\Delta} -1 \right]
\end{equation}
The probe limit result, in which the signal does not backreact on the geometry, can be obtained by assuming that $G_N p^+$ is small. Expanding $D(p^+)$ to first order in $G_N\,p^+$, we obtain
\begin{equation}
    C_\text{probe}=  c_{\Delta_{\phi}} e^{- \Delta_{\phi} \frac{2\pi}{\beta}(t_L+t_R)} \int_0^{\infty} dp^+ \,  (p^+)^{2\Delta_{\phi}-1} e^{- 4 i p^{+} \cosh  \frac{\pi}{\beta} (t_R-t_L) e^{-\frac{\pi}{\beta}(t_R+t_L)}}   e^{- i \delta X_{+} p^+ } \,.
\end{equation}
where 
\begin{equation}
  \delta X_+=g\, G_N \Delta \,\left[ 2 \cosh \left( \frac{2\pi}{\beta}t_0 \right) \right]^{-2\Delta-1}  \,.
\end{equation}
corresponds to the null shift introduced by the negative energy shock.
Performing the integral in $p^+$, we obtain
\begin{equation}
    C_\text{probe}= \left[ 2 \cosh \frac{\pi}{\beta}(t_R-t_L)+\frac{\delta X_+}{2} e^{\frac{\pi}{\beta}(t_R+t_L)}\right]^{-2\Delta_{\phi}}
\end{equation}
Introducing $U_0=e^{\frac{2\pi}{\beta}t_0}$, and using $\delta X^-=-\delta X_+$ we can write the shift as follows
\begin{equation} \label{eq:WormOpeningGJW}
  \delta X^-=- g\, G_N \Delta \left( \frac{U_0}{1+U_0^2}\right)^{2\Delta+1}\,.
\end{equation}
Note that $\delta X^- <0$ for $g>0$, corresponding to a traversable wormhole. See Figure~\ref{fig:info-tranfer}.
The dependence of $\delta X^-$ on $U_0$ perfectly matches the wormhole opening obtained by point-splitting for an instantaneous double-trace deformation in \cite{Freivogel:2019whb}. The dependence on the scaling dimension differs from \cite{Freivogel:2019whb} due to a different choice of normalization for boundary correlators.

The case where the signal backreacts on the geometry can be studied by numerically evaluating the integrals in (\ref{eq-Cnumerical}). However, one should note that the integral (\ref{eq-Cnumerical}) only converges for $\Delta<1/2$. The divergent behavior for $\Delta>1/2$ is due to the presence of high energy modes in the signal's wave function. This divergence can be eliminated by considering a smeared version of the signal operators, whose wave function does not contain high energy modes.

\subsection{Change of energy in GJW setup}
The change in the energy of the system due to the double trace deformation can be extracted from $\langle \psi(t)|H_r|\psi(t)\rangle$, where the perturbed state is given by
\be
|\psi(t) \rangle =e^{-H_0(t-t_0)} U(t,t_0) |\text{TFD} \rangle
\ee
where $U(t,t_0)=\mathcal{T}\,e^{-i \int_{t_0}^{t}dt_1\,\delta H(t_1)}$, where $\delta H(t_1) = g\, \delta(t_1-t_0) \,\mathcal{O}_L(-t_1)\,\mathcal{O}_R(t_1)$. At first order in $g$, we obtain
\begin{eqnarray}
\delta E_r &=& i \int_{t_0}^{t} dt_1 \, \delta(t_1-t_0) \langle \text{TFD} [\delta H(t_1),H_R]] |\text{TFD} \rangle \\
&=& g  \langle \text{TFD} |  \dot{\mathcal{O}}_R(t_0)\, \mathcal{O}_L(-t_0) |\text{TFD} \rangle\,.
\end{eqnarray}
The corresponding change of entropy can be obtained as $\delta S=\beta \, \delta E_R $.
For a thermal two-point function of the form 
\begin{equation}
    \langle \mathcal{O}_L(-t_0) \mathcal{O}_R(t_0)\rangle =\left[ \cosh \left( \frac{2\pi}{\beta}t_0 \right) \right]^{-2\Delta}\,,
\end{equation}
one obtains
\begin{equation}
    \delta E_r \sim g \,\frac{4\pi \Delta}{\beta} \,
    \frac{\sinh \!\left(\tfrac{2\pi}{\beta}t_0\right)}{\left[ \cosh \!\left(\tfrac{2\pi}{\beta}t_0\right) \right]^{2\Delta+1}}\,.
\end{equation}
Introducing the variable \(w_0 = e^{2\pi t_0/\beta}\), this can be written as
\begin{equation} \label{eq:deltaE-GJW}
    \delta E_r \sim g \,\frac{w_0 - w_0^{-1}}{\left( w_0 + w_0^{-1} \right)^{2\Delta+1}} \,.
\end{equation}
The corresponding change in entropy is then obtained from the first law of thermodynamics,
\begin{equation}\label{eq:deltaS-GJW}
    \delta S =\beta\, \delta E_r \sim  \beta \,g\frac{w_0 - w_0^{-1}}{\left( w_0 + w_0^{-1} \right)^{2\Delta+1}} \,.
\end{equation}

\section{Rindler coordinates}
\label{app:C}
In Rindler coordinates, the metric in the gravitating region and in the baths region takes the following form 
\be\label{eq:metric}
ds^2_{AdS} =-  \frac{4 \pi^2 \ell^2}{\beta^2} \frac{d y^+ dy^-}{\sinh^2{\frac{\pi}{\beta}(y^- - y^+)}}~, ~~ ds^2_{bath} = - \frac{\ell^2}{\epsilon_{UV}^2} dy^+ dy^-~.
\ee
where the AdS cutoff is $z = \epsilon_{UV}$, and the scale factor $1/\epsilon_{UV}^2$ guarantees that these two metrics agree at the cutoff. We parametrize points in the $y$-plane by
\be
y_L^\pm = t \mp z~, ~~ y_R^\pm = t \pm z~.
\ee
The dilaton has the profile
\be
\phi = \phi_0 + \frac{2 \pi \phi_r}{\beta} \frac{1}{\tanh \frac{\pi}{\beta}(y^- - y^+)}~.
\ee

\bibliographystyle{JHEP}
\bibliography{Main.bib}

\end{document}

%% file: macros.tex
\def\ie{i.e.~}

\def\ket#1{| #1 \rangle}

\newcommand{\cO}{{\mathcal{O}}}

\newcommand{\be}{\begin{equation}}
\newcommand{\ee}{\end{equation}}
\newcommand{\bea}{\begin{eqnarray}}
\newcommand{\eea}{\end{eqnarray}}
\newcommand{\lel}{\left\langle}
\newcommand{\rir}{\right\rangle}

\let\emptyset\varnothing